\documentclass[10pt,conference]{IEEEtran}
\IEEEoverridecommandlockouts
\usepackage{graphicx}
\usepackage{subcaption}
\usepackage{amsmath}
\usepackage{mathtools} 
\usepackage{cite}
\usepackage{amsmath,amssymb,amsfonts}
\usepackage{algorithmic}
\usepackage{textcomp}
\usepackage{xcolor}
\usepackage{algorithm}
\usepackage{setspace}
\usepackage{booktabs}
\usepackage{hyperref}
\usepackage{enumitem}
\usepackage{balance}
\setlist[enumerate]{leftmargin=*, label=(\arabic*), itemsep=0pt, topsep=0pt}
\setlist[itemize]{leftmargin=*, itemsep=0pt, topsep=0pt}
\usepackage{multirow}
\usepackage{tcolorbox}
\usepackage{bm}
\usepackage{pifont}
\usepackage{xcolor,colortbl}
\definecolor{c1}{cmyk}{0,0.6175,0.8848,0.1490}
\definecolor{c2}{cmyk}{0.1127,0.6690,0,0.4431}
\definecolor{c3}{cmyk}{0.3081,0,0.7209,0.3255}
\definecolor{c4}{cmyk}{0.6765,0.2017,0,0.0667}
\definecolor{c5}{cmyk}{0,0.8765,0.7099,0.3647}

\definecolor{lightgrey}{rgb}{0.93,0.93,0.93}

\newtcbox{\hlprimarytab}{on line, rounded corners, box align=base, colback=c3!10,colframe=white,size=fbox,arc=3pt, before upper=\strut, top=-2pt, bottom=-4pt, left=-2pt, right=-2pt, boxrule=0pt}
\newtcbox{\hlsecondarytab}{on line, box align=base, colback=red!10,colframe=white,size=fbox,arc=3pt, before upper=\strut, top=-2pt, bottom=-4pt, left=-2pt, right=-2pt, boxrule=0pt}

\newcommand{\uashifted}{\raisebox{0.5\depth}{\tiny$\uparrow$}}

\newcommand{\ua}[1]{{\footnotesize\hlprimarytab{\uashifted{#1}}}}
\def\BibTeX{{\rm B\kern-.05em{\sc i\kern-.025em b}\kern-.08em
    T\kern-.1667em\lower.7ex\hbox{E}\kern-.125emX}}
\newcommand{\boxmargin}{4pt}

\newcommand{\ourmodel}{\textsc{EfficientEdit}}

\newtcolorbox{RQbox}{
    colback=blue!10!white,  % 浅蓝色背景
    arc = 0pt, outer arc = 0pt,
    boxsep=0pt, left = 3pt, right = 0pt, top = 0pt, bottom = 0pt, 
    leftrule=3pt, bottomrule=0pt, toprule=0pt, rightrule=0pt,
    left = \boxmargin, right = \boxmargin, top = \boxmargin, bottom = \boxmargin
}

\usepackage{listings}
\usepackage{fancyvrb}
\definecolor{codegreen}{rgb}{0.0,0.5,0.0}
\definecolor{codegray}{rgb}{0.4,0.4,0.4}
\definecolor{codeblue}{rgb}{0.0,0.0,0.8}
\definecolor{backcolour}{rgb}{0.97,0.97,0.97}
\definecolor{highlightyellow}{rgb}{1,0.95,0.8}
\definecolor{instructionbg}{rgb}{0.94,0.96,1}
\definecolor{bordercolor}{rgb}{0.6,0.6,0.6}
\lstdefinestyle{pythonstyle}{
    backgroundcolor=\color{backcolour},   
    commentstyle=\color{codegray},
    keywordstyle=\color{codeblue},
    numberstyle=\tiny\color{codegray},
    stringstyle=\color{codegreen},
    basicstyle=\ttfamily\small,
    breakatwhitespace=false,         
    breaklines=true,                 
    captionpos=b,                    
    keepspaces=true,                 
    numbers=left,                    
    numbersep=3pt,                  
    showspaces=false,                
    showstringspaces=false,
    showtabs=false,                  
    tabsize=2,
    language=Python,
    xleftmargin=0.5em,
    framexleftmargin=0.5em
}
\usepackage{cleveref}
\crefname{figure}{Figure}{Fig.}
\crefname{table}{Table}{Tab.}
\crefname{section}{Section}{Sec.}
\crefname{equation}{Equation.}{Eq.}
\crefname{algorithm}{Algorithm}{Alg.}
\begin{document}

\title{
% arxiv %
\ourmodel{}: Accelerating Code Editing via Edit-Oriented Speculative Decoding

% \ourmodel{}: Accelerating Code Editing via Edit-Oriented Speculative Decoding
% \ourmodel{}: Accelerating Code Editing via Reuse and Generation
% FastEdit: Accelerating Code Editing via Reuse-Localize-Edit Oriented and Flexible Speculative Decoding
}
% \author{\IEEEauthorblockN{Anonymous Authors}}

\author{
    \IEEEauthorblockN{Peiding Wang$^{1}$, Li Zhang$^{1}$, Fang Liu$^{1*}$\thanks{* Corresponding author}, Yinghao Zhu$^{1}$, Wang Xu$^{3}$, Lin Shi$^{2*}$ \\
    Xiaoli Lian$^{1}$, Minxiao Li$^{1}$, Bo Shen$^{4}$, An Fu$^{4}$}
    \IEEEauthorblockA{$^{1}$State Key Laboratory of Complex \& Critical Software Environment, School of Computer Science and Engineering, \\ Beihang University, China}
    \IEEEauthorblockA{$^{2}$School of Software, Beihang University, China}

    \IEEEauthorblockA{$^{3}$Tsinghua University, China}
    \IEEEauthorblockA{$^{4}$Huawei Cloud Computing Technologies Co., Ltd., China}
    \IEEEauthorblockA{
    Email: \{\href{mailto:wangpeiding@buaa.edu.cn}{wangpeiding}, \href{mailto:fangliu@buaa.edu.cn}{fangliu},
    \href{mailto:shilin@buaa.edu.cn}{shilin}\}@buaa.edu.cn
    }
}

%     \IEEEauthorblockA{
%     \{\href{mailto:wangylin36@mail.sysu.edu.cn}{wangylin36}, \href{mailto:chenjch86@mail.sysu.edu.cn}{chenjch86},
%     \href{mailto:zhzibin@mail.sysu.edu.cn}{zhzibin}\}@mail.sysu.edu.cn,
%     \{\href{mailto:wangyli58@mail2.sysu.edu.cn}{wangyli58}, \href{mailto:guody5@mail2.sysu.edu.cn}{guody5}\}@mail2.sysu.edu.cn,
%     % \{\href{mailto:zhangruikai1@huawei.com}{zhangruikai1}, \href{mailto:mayuchi1@huawei.com}{mayuchi1}\}@huawei.com
%     }
%     \IEEEauthorblockA{$^{2}$ Huawei Cloud Computing Technologies Co., Ltd., Shenzhen, China}
%     \IEEEauthorblockA{
%     % \{\href{mailto:wangylin36@mail.sysu.edu.cn}{wangylin36}, \href{mailto:chenjch86@mail.sysu.edu.cn}{chenjch86},
%     % \href{mailto:zhzibin@mail.sysu.edu.cn}{zhzibin}\}@mail.sysu.edu.cn,
%     % \{\href{mailto:wangyli58@mail2.sysu.edu.cn}{wangyli58}, \href{mailto:guody5@mail2.sysu.edu.cn}{guody5}\}@mail2.sysu.edu.cn,
%     \{\href{mailto:zhangruikai1@huawei.com}{zhangruikai1}, \href{mailto:mayuchi1@huawei.com}{mayuchi1}\}@huawei.com
%     }
% }

\maketitle

\begin{abstract}
Large Language Models (LLMs) have demonstrated remarkable capabilities in code editing, substantially enhancing software development productivity. 
% However, a critical factor in practical deployment is LLMs' decoding speed.
% The decoding speed is important in code editing which influence the efficiency.
% However, most existing approaches typically employ end-to-end autoregressive generation by LLMs for code editing tasks,
However, the inherent complexity of code editing tasks forces existing approaches to rely on LLMs' autoregressive end-to-end generation, where decoding speed plays a critical role in efficiency.
% especially in long code context decoding latency becoming prohibitive, which substantially limits editing efficiency.
% The decoding speed plays a critical role in determining the efficiency of these models during code editing tasks.
While inference acceleration techniques like speculative decoding are applied to improve the decoding efficiency, these methods fail to account for the unique characteristics of code editing tasks, where changes are typically localized and existing code segments are reused. To address this limitation, we propose \ourmodel{}, a novel method that improves LLM-based code editing efficiency through two key mechanisms based on speculative decoding: (1) effective reuse of original code segments while identifying potential edit locations, and (2) efficient generation of edit content via high-quality drafts from edit-oriented draft models and a dynamic verification mechanism that balances quality and acceleration. 
% By adopting this iterative reuse-and-generation approach, \ourmodel{} achieves notable efficiency improvements without compromising output quality. 
Experimental results show that \ourmodel{} can achieve up to 10.38× and 13.09× speedup compared to standard autoregressive decoding in CanItEdit and CodeIF-Bench, respectively, outperforming state-of-the-art inference acceleration approaches by up to 90.6\%. The code and data are available at \url{https://github.com/zhu-zhu-ding/EfficientEdit}.

\end{abstract}


\section{Introduction}

In recent years, Large Language Models (LLMs) (e.g., GPT-4 \cite{openai2024gpt}, Claude \cite{claude}, DeepSeek-Coder \cite{zhu2024deepseek}, Qwen-Coder \cite{qwencoder}, CodeLlama \cite{roziere2023codellama}) have demonstrated remarkable success in code editing related tasks \cite{canitedit,aggarwal2025nextcoder,muennighoff2024octopack}. Concurrently, AI-powered development tools such as GitHub Copilot \cite{GitHub-Copilot} and Cursor \cite{Cursor} are significantly transforming modern software development workflows by applying LLMs to code editing. Empirical studies also reveal that code editing—which includes modifying \cite{instructcoder,li2023codeeditor}, refactoring \cite{guo2024codeeditorbench}, or debugging existing code \cite{just2014defects4j}—constitutes a highly frequent activity during the lifecycle of a software project \cite{kitchin2014code,mozannar2024reading,Chen2025Deep}. Beyond the correctness of code edits, generation efficiency represents another critical factor that affects overall development productivity.
% Beyond code correctness, the decoding efficiency of LLMs critically impacts developer productivity through generation latency.

\begin{figure}[t!]
    \centering
    \includegraphics[width=1.0\linewidth]{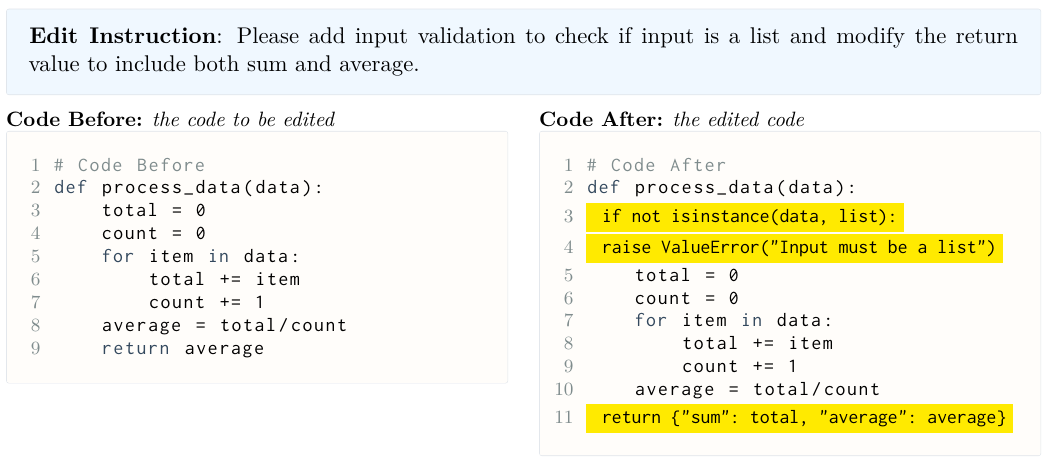}
    \caption{A motivating example of code editing. 
    % The instruction requires adding input validation and modifying the return value. 
    Highlighted contents in ``Code After'' indicate newly generated or modified code. A significant portion of ``Code After'' (approximately 70\%) is reused from ``Code Before''. 
    % underscoring the potential for reuse-centric acceleration.
    }
    \label{fig:motivation_example}
    \vspace{-0.6cm}
\end{figure}
% The code editing can be catetoried into xxx and xxx. end 
Accurately inferring precise edit locations from natural language instructions poses a significant challenge in code editing. As illustrated in \Cref{fig:motivation_example}, the instruction ``\textit{Please add input validation to check if input is a list and modify the return value to include both sum and average}'' specifies what to change but not necessarily where to insert new code (\textit{e.g.}, the exact line for validation checks) or how to integrate changes with existing code structures (\textit{e.g.}, restructuring the return statement). This ambiguity often necessitates complex reasoning from the LLM. Consequently, most existing code editing approaches employ end-to-end generation paradigm using LLMs \cite{lin2023cct5,zhang2022coditt5,wei2024coeditor,canitedit,muennighoff2024octopack,aggarwal2025nextcoder,lozhkov2024starcoder}, which predominantly rely on autoregressive decoding to produce outputs token by token. This paradigm significantly limits the efficiency of LLM-based code editing. For instance, editing a 1024-token code snippet with DeepSeek-Coder-33B \cite{zhu2024deepseek} requires approximately 86 seconds on 4 NVIDIA GeForce RTX 4090 GPUs. 

To improve LLM's inference efficiency, speculative decoding \cite{leviathan2023fastinferencetransformersspeculative,chen2023acceleratinglargelanguagemodel} offers a promising solution without modifying the underlying LLM architecture. This approach employs a small draft model to rapidly generate candidate output tokens (drafts), which are then verified by the target LLM in a single forward pass to ensure drafts strictly matching the target LLM's output. Building upon the speculative decoding paradigm, various approaches are proposed to accelerate general-purpose text generation tasks \cite{zhao2024ouroboros,lookahead,zhang2024draft}. Recent work has further explored replacing parameterized draft models with retrieval systems to simplify draft model selection \cite{rest}. In particular, FastFixer\cite{fastfix} first applied speculative decoding to APR (Automatic Program Repair) tasks, substantially improving program repair efficiency. 
Most existing inference acceleration methods achieve $2\sim3\times$ speedup in inference, effectively improving LLMs' inference speed. 

Although these efforts have shown promising performance in accelerating LLM inference, we have identified the following problems in code editing:
\begin{itemize}
    \item \textbf{Redundant Code Generation.} End-to-end code editing with LLMs often produces redundant code due to the autoregressive decoding mechanism \cite{xu2022learning,dong2025rethinking}, leading to inefficient inference and unnecessary computational overhead. As shown in Figure \ref{fig:motivation_example}, the generated output contains approximately 70\% redundant code segments.
    \item \textbf{Limitations of Speculative Decoding.} Limited by the ordinary quality of the draft model and the strict verification process, the benefits of using speculative decoding in code editing scenarios are not significant. It still takes approximately 30-40 seconds to generate an edited code of 1024 tokens for DeepSeek-Coder-33B when running on 4 NVIDIA GeForce RTX 4090 GPUs. 
\end{itemize}

To address these limitations, we propose \ourmodel{}, an efficient edit approach for code editing via speculative decoding. As indicated in \Cref{fig:motivation_example}, to leverage the abundant redundant fragments in the code to be edited, we employ speculative decoding to efficiently reuse these fragments while identifying potential edit locations through a single forward inference of the target model. Once potential edit positions are determined, we enhance the target model's editing efficiency using high-quality drafts generated by a draft model fine-tuned with our proposed edit-content-oriented strategy. Moreover, we introduce an effective verification mechanism that optimizes the trade-off between generation quality and efficiency, further improving the target model's performance in generating edit content. Extensive evaluation across diverse benchmarks, including crowd-sourced edits (CanItEdit \cite{canitedit}) and interactive editing (CodeIF-Bench \cite{CodeIF-Bench}) spanning function-, class-, and repository-level tasks, demonstrates substantial inference speedup, achieving up to 8.26$\times$ for Qwen-32B-Instruct and 13.09$\times$ for DeepSeek-Coder-33B-Instruct, significantly outperforming baseline acceleration methods.

Our key contributions are summarized as follows:
\begin{itemize}
    \item We propose \ourmodel{}, an efficient approach for improving LLM's code editing efficiency by synergistically combining code reuse with accelerated generation of new content.
    \item We propose the edit-oriented draft models that improves the target LLMs' code editing efficiency in speculative decoding.
    % 改成model
    \item We conduct a comprehensive evaluation of \ourmodel{} and demonstrate that it achieves state-of-the-art code editing efficiency, while maintaining or even improving edit quality.
    % \item We conduct a comprehensive evaluation of \ourmodel{} and several state-of-the-art baselines on diverse code editing scenarios, demonstrating speedups of up to 8.26$\times$ for Qwen2.5-Coder-32B-Instruct and 13.09$\times$ for DeepSeek-Coder-33B-Instruct, while maintaining or even improving edit quality.
    % \item We design an edit-oriented fine-tuning method to obtain a high-quality draft model and an entropy-aware dynamic verification mechanism that balances generation efficiency and quality to improve the efficiency of edits generation with the target LLMs.
\end{itemize}

\section{Preliminaries}
% \lin{add something here to explain why you need introduce the two decodings}
\subsection{Autoregressive Decoding}
% \fang{Most LLMs utilize an autoregressive decoding process, generating output tokens $t_i$ sequentially based on the input $prefix$ and previously generated tokens $t_1, ..., t_{i-1}$: 
% \begin{equation}
%     (p_i,t_i) = \texttt{LLM}(t_1, ..., t_{i-1};\text{prefix}) 
% \end{equation}}

Autoregressive decoding is a prevalent approach in LLM inference, wherein output tokens are sequentially generated through progressive forward predictions.

Formally, given the LLM $\texttt{LLM}$ and the prefix token sequence $(t_1, ..., t_{n-1})$ and a new input token $t_{n}$, the model predicts the probability distribution of the subsequent token as follows:

\begin{equation}
\bm{p}_{n+1} = \texttt{LLM}(t_n;t_1, ..., t_{n-1})
\end{equation}

The newly generated token $t_{n+1}$ is either selected as the top-1 prediction from $\bm{p}_{n+1}$ or probabilistically sampled according to the distribution $\bm{p}_{n+1}$.
Then the new selected token $t_{n+1}$ is input to the LLM to generate the next token.

Inherently, the token-by-token generation mechanism of autoregressive decoding introduces computational inefficiencies, leading to a linear escalation of inference latency that is proportional to both the generated sequence length and the model's architectural complexity.

\subsection{Speculative Decoding}
To enhance decoding efficiency, Speculative decoding has been proposed in recent literature~\cite{leviathan2023fastinferencetransformersspeculative, chen2023acceleratinglargelanguagemodel}. This innovative approach accelerates autoregressive generation through a draft-and-verify paradigm. 
During the draft stage, these methods employ a fast-inference draft model ($\texttt{LLM}_{\text{draft}}$) to generate $\gamma$ candidate tokens (drafts). 
These candidate tokens are subsequently verified in parallel via a forward inference through the target Large Language Model ($\texttt{LLM}_{\text{target}}$), thereby significantly improving inference speed.

During the draft stage, a draft model generates $M$ draft tokens $(d_1,\cdots,d_{M})$ autoregressively given the prefix token sequence $(t_1,t_2,\cdots,t_{n})$.
\begin{equation}
    \bm{q}_{M+1} = \texttt{LLM}_{\text{draft}}(d_{M}; t_1, ..., t_{n},d_1,\cdots,d_{M-1}) 
\end{equation}
Token $d_{M+1}$ is sampled from the distribution of $\bm{q}_{M+1}$ (but usually greedy).

During the verify stage, the $M$ draft tokens $(d_1,\cdots,d_{M})$ are input to the target language model in a forward pass as follows:
\begin{equation}
    \bm{p}_{1},\cdots,\bm{p}_{M} = \texttt{LLM}_{\text{target}}(d_1,\cdots,d_{M}; t_1, ..., t_{n}) 
\end{equation}
Speculative decoding is used to determine which token would be accepted as follows:
\begin{equation}
\begin{split}
    \epsilon_i < \frac{\bm{p}_i[d_i]}{\bm{q}_i[d_i]}
\end{split}
\end{equation}
where $\epsilon_i \sim \mathcal{U}\left([0,1]\right)$.
$\bm{p}_i[d_i]$ and $\bm{q}_i[d_i]$represents the token $d_i$ corresponding probability in $\bm{q}_i$ and $\bm{p}_i$. 
$\epsilon_i$ is a given hyperparameter.
Intuitively, the draft token can be accepted if it has a higher probability in the target model's distribution. For lower-probability draft tokens, the decision is made stochastically based on the probability difference. 

When the $\texttt{LLM}_{\text{target}}$ rejects the draft token $d_i$, the above process is repeated by resampling a token $t_i$ from the target model's probability vector $\bm{p}_i$. 
In the ideal case, all $M$ tokens generated by $\texttt{LLM}_{\text{draft}}$ are accepted by $\texttt{LLM}_{\text{target}}$ through speculative decoding. 
Moreover, a new token $t_{M+1}$ would be sampled from the probability distribution $\bm{p}_{M}$.
Thus $M+1$ tokens are accepted by a single forward of the target model, thereby significantly improving its inference speed.
\subsection{Task Definition}
% \lin{Typically task definition is introduced before Methodology, better place in II.}
This study investigates end-to-end code editing tasks where edit positions cannot always be accurately determined from natural language specifications alone. 
% Given the inherent difficulty in determining optimal edit locations within source code, 
We formulate the code editing task as an end-to-end sequence transformation problem:

\begin{equation}
    c' = E(c,I) 
\end{equation}
where $c$ is the original input code to be modified, $I$ denotes the edit instruction (requirement), and $c'$ is the output code after applying the edit. Each line of the output code $l_i \in c'$ can be further categorized into two subsets:
\begin{itemize}
    \item Reused lines from the input code: $\mathcal{R}=\{l_i | l_i \in c \cap c' \}$
    \item Newly generated lines: $\mathcal{G}=\{l_i \in c' \setminus c \}$
\end{itemize}
Thus, the output code $c'$ is formally represented as:
\begin{equation}
    c'=\{ l_1, l_2, \dots, l_n \}, \ \text{where} \ l_i \in (\mathcal{R} \cup \mathcal{G})
\end{equation}

\section{Methodology}
\begin{figure*}
	\centering
	\includegraphics[width=1\textwidth]{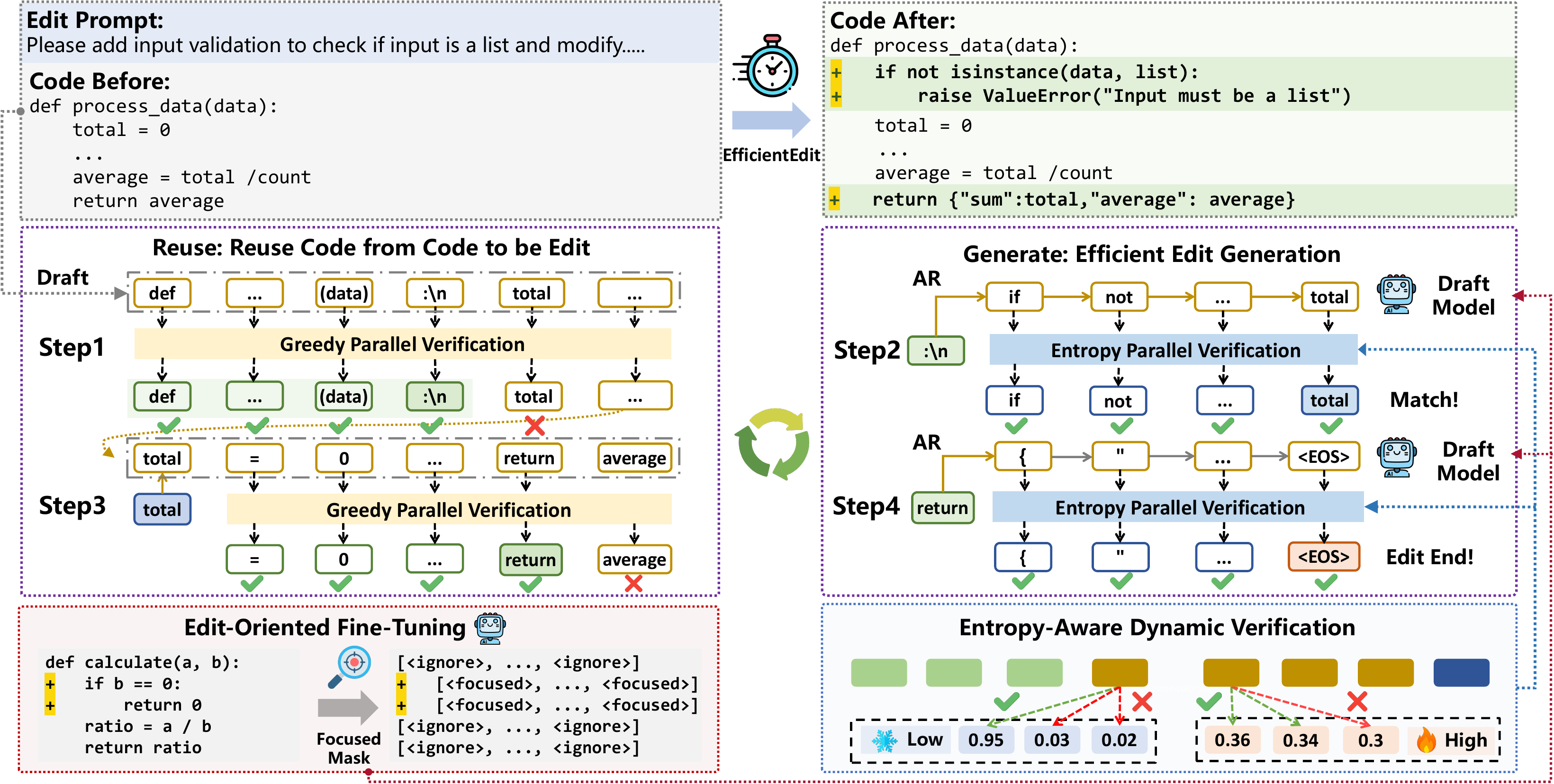}
	\caption{The overview of \ourmodel{}. }
	\label{fig:total_process}
        \vspace{-15pt}
\end{figure*}

In this section, we present our proposed approach, \ourmodel{}, for enhancing LLM-based code editing efficiency through speculative decoding. As shown in Figure \ref{fig:total_process}, \ourmodel{} operates in two key phases \textit{Reuse} and \textit{Edit}. \textit{Reuse} treats the original code (or the remaining code) as the high-quality draft and performs parallel verification via a single forward pass of the target LLM, enabling efficient reuse of unmodified code segments. The draft segments rejected by the target LLM are preserved as candidate drafts for subsequent processing. Then, \textit{Edit} uses the rejection points identified by the target LLM as starting positions for generating edited content. This process is further optimized through an edit-oriented draft model and a novel entropy-aware verification mechanism. After completing the edited content, a prefix matching algorithm is used to continue reusing code segments from the candidate drafts. This iterative reuse-generate paradigm significantly accelerates LLM-based code editing while maintaining output quality.
% \ourmodel{} comprises two key components: 
% \begin{itemize}
%     \item \textbf{Locate: Self-Editing Location with Code Reuse}. The \textit{Locate} component achieves lightweight edit location through efficient reuse of $\mathcal{R}$. 
%     \item \textbf{Edit: Efficient Edit Generation}. The \textit{Edit} component efficiently generates edit content using a high-quality draft model and dynamic verification of drafts.
% \end{itemize}
% \subsection{Task Definition}
% \lin{Typically task definition is introduced before Methodology, better place in II.}
% This study investigates end-to-end code editing tasks where edit positions cannot always be accurately determined from natural language specifications alone. 
% % Given the inherent difficulty in determining optimal edit locations within source code, 
% We formulate the code editing task as an end-to-end sequence transformation problem:

% \begin{equation}
%     c' = E(c,I) 
% \end{equation}
% where $c$ is the original input code to be modified, $I$ denotes the edit instruction (requirement), and $c'$ is the output code after applying the edit. Each line of the output code $l_i \in c'$ can be further categorized into two subsets:
% \begin{itemize}
%     \item Reused lines from the input code: $\mathcal{R}=\{l_i | l_i \in c \cap c' \}$
%     \item Newly generated lines: $\mathcal{G}=\{l_i \in c' \setminus c \}$
% \end{itemize}
% Thus, the output code $c'$ is formally represented as:
% \begin{equation}
%     c'=\{ l_1, l_2, \dots, l_n \}, \text{where} \quad l_i \in (\mathcal{R} \cup \mathcal{G})
% \end{equation}

\subsection{Reuse: Reuse Code from Code to be Edit}

\ourmodel{} leverages $c$ as high-quality drafts for accelerated reuse of $\mathcal{R}$ and identifies editing positions for newly generated code $\mathcal{G}$. Specifically, since there is usually a large amount of redundant code segments in the code to be edited $c$ and the edited code $c'$, we employ speculative decoding to parallelize and reuse these repetitive code segments via the $\texttt{LLM}_{\text{target}}$. For speculative decoding, the selection of $\gamma$ is import and challenge. While a larger $\gamma$ can reduce the number of costly $\texttt{LLM}_{\text{target}}$ inferences when draft quality is high, excessive candidate token rejections may conversely increase verification overhead and draft generation time. 

In code editing, we exploit the structural properties of code: $c'$ often reuses complete lines or segments from $c$. 
Furthermore, $c$, as a natural draft, does not require any additional generation time. This observation motivates our approach of directly using $c$ (or its remaining portions) as the draft to reuse these multi-token code segments and significantly reduce inference costs. This approach achieves a key advantage: \textbf{there is no need to determine an optimal draft length $\gamma$ for reuse, as a single forward pass by $\texttt{LLM}_{\text{target}}$ can verify a long sequence of tokens from $c$.} After $\texttt{LLM}_{\text{target}}$ accepts reusable code segments in a single forward inference, intuitively, token rejection by $\texttt{LLM}_{\text{target}}$ implicitly identifies the potential starting position of a segment in $\mathcal{G}$, effectively pinpointing an edit location without additional computational cost. 
At this point, \ourmodel{} transitions to the \textit{Generate} phase to efficiently generate $\mathcal{G}$.
When in the \textit{Generate} phase and needing to switch back to \textit{Reuse}, we employ a prefix matching algorithm to identify prefixes of the remaining $c$ that match the suffix of the currently generated $\mathcal{G}$, using these match points as new draft starting points. The subsequent tokens from these positions in $c$ then serve as draft candidates for parallel verification by $\texttt{LLM}_{\text{target}}$. Another key advantage of \textit{Reuse} is that \textbf{it reuses code while also achieving self-editing positioning.}

Given the current generated prefix $prefix$ and the original code $c$ (or its unverified remainder), the \textit{Reuse} phase uses a portion of $c$ as a draft sequence $draft = (d_1, d_2, \dots, d_m)$. This $draft$ is appended to $prefix$. A single forward pass of $\texttt{LLM}_{\text{target}}$ over $(prefix, draft)$ yields a sequence of next-token probability distributions $(\bm{p}_1, \bm{p}_2, \dots, \bm{p}_m)$, where $\bm{p}_i$ is the distribution for the token in $draft$. Now, the verification method is as follows:
% The validation function determines how much of $draft$ is accepted is:
\begin{equation}
 \operatorname{decode}(\bm{p}_i) =  d_i \;\;\; {\forall i \leq j} 
% \text{Accept } d_i \text{ if all previous tokens are accepted and } \operatorname{decode}(\bm{p}_j) =  d_j
\label{eq:piecewise}
\end{equation}
where the accepted length of $draft$ is the maximum of $j$ satisfied above equation.  
% {\scriptsize
% \begin{equation}
% \begin{aligned}
% \begin{cases}
%     (prefix, d_1, \dots, d_m) & \text{if } \operatorname{decode}(\bm{p}_i) = d_i \text{ for all } i \in \{1, \dots, m\} \\
%     (prefix, d_1, \dots, d_{j-1}) & \text{if } \operatorname{decode}(\bm{p}_j) \neq d_j \text{ and } \operatorname{decode}(\bm{p}_i) = d_i \text{ for } i < j
% \end{cases}
% \end{aligned}
% \label{eq:piecewise}
% \end{equation}
% }
Effectively, tokens from $draft$ are accepted as long as they match the greedy decoding output of $\texttt{LLM}_{\text{target}}$ at each position. If a mismatch occurs at token $d_j$, all preceding accepted tokens $(d_1, \dots, d_{j})$ extend $prefix$ and switch to the \textit{Generate}. At the same time, we treat the remaining drafts $(d_{j+1}, \dots, d_{m})$ as reusable candidate drafts. In this paper, the $\operatorname{decode}$ function in \Cref{eq:piecewise} uses greedy decoding, i.e., the token with the highest probability is selected.

\subsection{Generate: Efficient Edit Generation}
When editing large amounts of content, the sequential nature of autoregressive decoding emerges as a critical efficiency bottleneck. And the inherent flexibility and complexity of code editing tasks pose significant challenges for obtaining and reusing high-quality drafts at edit locations similar to \textit{Reuse}. To address this, we adopt a speculative decoding paradigm that leverages both target and draft models to accelerate the generation of edited content.
% 自回归 edit

% 解释也用投机加速，需要编辑的部分代码大部分变化比较大，基于大模型/小模型 但是和上面draft不一样

% 基于insigt 提出两种方案

\ourmodel{} efficiently generates the edit content $\mathcal{G}$ by employing an $\texttt{LLM}_{\text{draft}}$, fine-tuned via \textit{Edit-Oriented Fine-Tuning}, to generate high-quality drafts that expedite $\texttt{LLM}_{\text{target}}$'s inference. Additionally, we introduce an \textit{Entropy-Aware Dynamic Verification} mechanism to balance generation quality and the acceleration effect for $\texttt{LLM}_{\text{target}}$.
The key insight is that \textbf{the $\texttt{LLM}_{\text{draft}}$ primarily needs to generate contextually appropriate drafts at $\texttt{LLM}_{\text{target}}$'s edit positions, and does not require strict greedy decoding-based verification across all tokens to achieve generation quality comparable to that of $\texttt{LLM}_{\text{target}}$.}

\textbf{Edit-Oriented Fine-Tuning.} 
% In speculative decoding, the $\texttt{LLM}_{\text{draft}}$'s ability to generate high-quality draft sequences critically impacts overall performance. 
In \ourmodel{}, $\texttt{LLM}_{\text{draft}}$ must produce high-quality drafts to accelerate $\texttt{LLM}_{\text{target}}$'s generation of $\mathcal{G}$. However, despite potential distributional similarity between $\texttt{LLM}_{\text{draft}}$ and $\texttt{LLM}_{\text{target}}$, it is usually difficult for $\texttt{LLM}_{\text{draft}}$ to generate token sequences that perfectly match $\texttt{LLM}_{\text{target}}$'s output. This difficulty can increase the verification cost for $\texttt{LLM}_{\text{target}}$, thereby affecting the acceleration. 
To address this issue, we propose a training strategy focused on edit positions. First, we use $\texttt{LLM}_{\text{target}}$ to generate $c'$ from $I$ and $c$. Next, we identify lines in $c'$ that do not exist in $c$ and label them as $\mathcal{G}$. By masking all $\mathcal{R}$ tokens in $c'$ except those belonging to $\mathcal{G}$, we exclude $\mathcal{R}$ from loss computation during the fine-tuning of $\texttt{LLM}_{\text{draft}}$. This design is motivated by the intuition that $\texttt{LLM}_{\text{draft}}$ does not need to learn how to perfectly generate $\mathcal{R}$ or locate the editing position for $\texttt{LLM}_{\text{target}}$; instead, it should focus solely on accurately predicting $\mathcal{G}$. Based on this, we design the following loss-masking training strategy:
\begin{equation}\label{eq:masked-loss}
\mathcal{L}_{\text{EoFT}} = \sum_{i \in \mathcal{G}_{\text{tokens}}} \mathcal{L}_{\text{origin}}(y_i|x_i)
\end{equation}
where $\mathcal{L}_{\text{origin}}$ is the cross-entropy loss, $y_i$ denotes the $\texttt{LLM}_{\text{target}}$ label of the $i$-th token within the tokenized representation of $\mathcal{G}$ (denoted $\mathcal{G}_{\text{tokens}}$), and $x_i$ is the corresponding token generated by $\texttt{LLM}_{\text{draft}}$.

\textbf{Entropy-Aware Dynamic Verification.} Another critical factor influencing the performance of draft models in enhancing the inference performance of $\texttt{LLM}_{\text{target}}$ is the validation process. On one hand, since draft models are hard to generate outputs identical to $\texttt{LLM}_{\text{target}}$, an overly strict validation mechanism reduce speedup gains and reject correct content generated by $\texttt{LLM}_{\text{draft}}$. On the other hand, relaxing the validation threshold may accept the incorrect code so compromise output quality. 
As shown in Figure \ref{fig:total_process}, the LLM's probability distribution at each position can exhibit substantial variation. When the highest-probability token dominates the distribution with a significantly greater probability than other tokens, this reflects high confidence in the target LLM's inference; conversely, when the top-1 token's probability margin over alternatives is narrow, it indicates low confidence. Inspired by this observation, we propose an adaptive verification mechanism: for positions where the $\texttt{LLM}_{\text{target}}$ demonstrates high confidence (characterized by a large probability gap between the top token and others), the $\texttt{LLM}_{\text{draft}}$'s output must exactly match the target LLM's prediction, while for low-confidence positions, we relax the verification threshold to accept the draft output if it matches any of the target LLM's $top_\tau$ candidate tokens.
Specifically, given the base threshold hyperparameter $k$, at each token validation step we select the first $top_k$ probability vectors from $\texttt{LLM}_{\text{target}}$ at this token and compute their entropy $H_{\text{norm}}$ using the following formula:
\begin{equation}
H_{\text{norm}} = \frac{-\sum \bm{p_i} \log \bm{p_i}}{\log top_k}, \ i\in\{1, \dots,top_k\}
\label{eq:entropy}
\end{equation}
$top_\tau$ represents the threshold for draft tokens generated by the $\texttt{LLM}_{\text{draft}}$ to be accepted by the $\texttt{LLM}_{\text{target}}$. Specifically, the draft token is accepted when the highest probability token of $\texttt{LLM}_{\text{draft}}$ is in the first $top_\tau$ positions in the distribution of $\texttt{LLM}_{\text{target}}$. 
\begin{equation}
\quad top_\tau =  \\
\begin{cases}
    \lceil k \cdot H_{\text{norm}}\rceil & \text{if }\lceil k \cdot H_{\text{norm}}\rceil \geq 1 \\
    1 & \text{if }\lceil k \cdot H_{\text{norm}}\rceil < 1\\
\end{cases}
\end{equation}
% \begin{equation}
% \begin{aligned}
% \mathrm{Judge}(t) = 
% \begin{cases}
%    True & \text{if }t_{1}\text{ in }(t'_1,\dots,t'_{top_\tau}) \\
%     False & \text{if }t_{1}\text{ not in }(t'_1,\dots,t'_{top_\tau})
% \end{cases}
% \end{aligned}
% \label{eq:piecewise}
% \end{equation}
Higher entropy values indicate greater uncertainty in the $\texttt{LLM}_{\text{target}}$ of the current location, which leads to a larger threshold $top_\tau$. 
% This means that when the uncertainty of $\texttt{LLM}_{\text{target}}$ at the current location is greater, the probability of adopting the deterministic labeling of $\texttt{LLM}_{\text{draft}}$ is greater. Specifically, labeling is accepted when the highest probability of labeling of $\texttt{LLM}_{\text{draft}}$ is in the first $top_\tau$ positions in the distribution of $\texttt{LLM}_{\text{target}}$. 
Conversely, the lower the entropy value, the lower the uncertainty of $\texttt{LLM}_{\text{target}}$, leading to a smaller threshold $top_\tau$. 
% This means that as the uncertainty of the $\texttt{LLM}_{\text{target}}$ at the current location gets smaller, the whiskers of the deterministic labeling of the draft model agrees with the deterministic labeling of the target model. 
In particular, $top_\tau$ takes 1 when the threshold is less than 1, i.e., $\texttt{LLM}_{\text{draft}}$ and $\texttt{LLM}_{\text{target}}$ must be strictly consistent. This validation method balances the acceptance rate and quality of high-entropy locations by adaptively adjusting $top_\tau$, while maintaining strict criteria for low-entropy locations. 

\section{Experimental Setups}

\subsection{Benchmarks}

To comprehensively evaluate \ourmodel{} on diverse code editing tasks, we employ the crowd-sourced benchmark CanItEdit \cite{canitedit} and the dialogue-based iterative editing benchmark CodeIF-Bench \cite{CodeIF-Bench}.
\begin{itemize}
    \item \textbf{CanItEdit} \cite{canitedit} contains 105 crowd-sourced Python code editing tasks, each with: (1) input/output code snippets, (2) two natural language instructions (descriptive and lazy), and (3) a hidden test suite. The dataset covers diverse domains (data structures, algorithms, NLP, etc.) and requires familiarity with Python libraries like NumPy and PyTorch. Tasks are categorized as corrective, refinement, or adaptive.
    \item \textbf{CodeIF-Bench} \cite{CodeIF-Bench} evaluates interactive code generation through multi-round dialogues, each dialogue set comprises an initial programming task (first round) followed by multiple independent editing instructions. It classifies tasks into three levels: L-1 (stand-alone), L-2 (intra-file context), and L-3 (cross-file context). We focus on L-1/L-2 across 9 real-world software instruction types, spanning algorithmic and repository-level tasks. L-3 is excluded due to context length and resource constraints. In this study, we exclusively evaluate the 2-Round scenario, where the first-round output serves as input code for subsequent editing, yielding a total of 762 editing problems with test cases.
\end{itemize}

\subsection{Training Dataset}

We select \textbf{InstructCoder} \cite{instructcoder} as the initial training dataset to obtain the high-quality draft model. InstructCoder is an instruction-tuning dataset for adapting LLMs to code editing tasks, covering repairs, refactoring, and transformations. Using self-instruct \cite{self-instruct} on GitHub-derived seed data, it generates 110K+ high-quality $(I,c,c')$ triples spanning diverse editing scenarios, originally generated by GPT-3.5 \cite{brown2020languagemodelsfewshotlearners}.

\subsection{Metrics}

\subsubsection{Edit Capability}

We use the \textbf{Pass@K} \cite{chen2021evaluatinglargelanguagemodels} metric to measure the LLMs' code edit capabilities. This metric is widely used in code generation and editing tasks \cite{canitedit,muennighoff2024octopack,chen2021evaluatinglargelanguagemodels}.
It reflects the probability that the model successfully generates functionally correct code at least once in $K$ attempts, computed as: 
\begin{equation}
\text{Pass@K} = 1 - \frac{\binom{n - c}{K}}{\binom{n}{K}}    
\end{equation}
where $n$ is the total number of generated code samples, $c$ is the number of correct solutions among these samples, and $K$ is the number of allowed attempts. In this paper, we primarily report Pass@1.

\subsubsection{Inference Efficiency}

Following existing work \cite{fastfix,zhao2024ouroboros,rest}, we use the following metrics to measure LLM inference efficiency for code editing:
\begin{itemize}
    \item \textbf{Tokens/s}: This metric represents the average number of tokens generated by LLMs per second during inference without mini-batching.
    \item \textbf{Speedup}: This metric measures the average speedup factor in tokens generated per second when using an inference acceleration method versus standard autoregressive inference.
\end{itemize}

\subsection{Baselines}

To assess the effectiveness of \ourmodel{}, we compare it with the following baselines: autoregressive inference (AR), standard speculative decoding (SD) \cite{leviathan2023fastinferencetransformersspeculative}, the state-of-the-art method Ouroboros \cite{zhao2024ouroboros} in NLP, and FastFixer \cite{fastfix} in program repair.

\begin{itemize}
    \item \textbf{Autoregressive Decoding (AR)}: AR is the fundamental token-by-token decoding approach employed by existing LLMs. In our evaluation, the autoregressive implementations of all methods use KV-Cache, which improves inference efficiency by memorizing key-value pairs in attention layers.
    \item \textbf{Speculative Decoding (SD)}: In this work, we adopt the standard speculative decoding framework where a smaller draft model generates candidate tokens, which are then verified in parallel by a larger target model. Note we use greedy decoding for the verification process in this baseline.
    \item \textbf{FastFixer}: FastFixer is a speculative decoding-based inference acceleration framework specifically designed for program repair. The method achieves substantial speedup by searching and reusing code segments as drafts from historical buggy code. We adapt the original FastFixer by replacing historical buggy code with the code that needs to be edited for general code editing tasks.
    \item \textbf{Ouroboros}: Ouroboros enhances draft model generation through phrases. The method employs heuristic techniques to extract reusable phrases from both rejected draft tokens and historical conversation contexts, thereby accelerating generation and improving speculative decoding performance. We adapt Ouroboros's phrase extraction approach from historical contexts to code editing, establishing it as an effective baseline.
\end{itemize}

\subsection{Implementation Details}

\textbf{Models}: We select the following model combinations to demonstrate the effectiveness of \ourmodel{}: 
\begin{itemize}
    \item \textbf{Qwen2.5-Coder Configuration:} We employ Qwen2.5-Coder-32B-Instruct as the target model combined with Qwen2.5-Coder-7B-Base as the draft model.
    
    \item \textbf{DeepSeek-Coder Configuration:} We employ DeepSeek-Coder-33B-Instruct as the target model combined with DeepSeek-Coder-6.7B-Base as the draft model.
    % \item Qwen2.5-Coder model combination: target model Qwen2.5-Coder-32B-Instruct and draft model Qwen2.5-Coder-7B-Base;
    % \item DeepSeek-Coder model combination: target model DeepSeek-Coder-33B-Instruct and draft model DeepSeek-Coder-6.7B-Base.
\end{itemize}
To avoid potential knowledge conflicts\cite{aggarwal2025nextcoder} with instruction-tuning data in our fine-tuning phase, we select the base models (Qwen2.5-Coder-7B-Base and DeepSeek-Coder-6.7B-Base) as the draft models and fine-tune them as the draft models in \ourmodel{}.

\textbf{Draft Model Training Data Preparation}: We randomly select 20K samples from the InstructCoder dataset and regenerate responses for each using DeepSeek-V3 (as DeepSeek-Coder-33B-Instruct API was unavailable) and Qwen-32B-Instruct. We use a rule-based method to distinguish between edit content $\mathcal{G}$ and reused code $\mathcal{R}$ in the edited code to apply our masked loss for fine-tuning the draft models. 

\textbf{Training Configuration}: Fine-tuning is performed using LoRA on 2 × NVIDIA RTX 4090 GPUs with hyperparameters: the maximum sequence length of 4096, the learning rate of 2e-5, the LoRA rank of 16, and the batch size of 32.

\textbf{Hardware and Implementation}: All experiments are performed on 5 × NVIDIA RTX 4090 GPUs, allocating 4 GPUs for the target model and 1 GPU for the draft model. The CPU used is an Intel(R) Xeon(R) Gold 6330 CPU. For the Qwen2.5-Coder configuration, the base entropy threshold $k$ is set to 3; for the DeepSeek-Coder configuration, $k$ is set to 5. Except for \ourmodel{}'s entropy-aware verification of edit content, all other speculative decoding methods (including the reuse phase of \ourmodel{}) use greedy decoding for verification. 
The hyperparameter draft length $\gamma$ for all evaluated methods need to set is 7.
% All evaluated methods need set the draft length γ=7 as the hyperparameter setting.
% The hyperparameter $\gamma$ for the speculative decoding is 7.

\section{Experimental Results}

We aim to address the following four research questions:

\begin{itemize}
    \item \textbf{RQ1:} How does \ourmodel{} perform on code editing? This question evaluates the overall effectiveness of \ourmodel{} in terms of both inference efficiency and edit quality compared to baselines.
    \item \textbf{RQ2:} How does each component of \ourmodel{} contribute to its performance? This question investigates the individual impact of \ourmodel{}'s core components.
    \item \textbf{RQ3:} How effective is Entropy-Aware Dynamic Verification compared to other  verification methods? This question compares our verification mechanism against other strategies.
    \item \textbf{RQ4:} How does \ourmodel{} perform in code editing tasks with different code reuse rates? This question examines the performance of \ourmodel{} across scenarios with varying proportions of reused versus edit content.
\end{itemize}

% \textbf{RQ1:} How does \ourmodel{} perform on code editing tasks? This question evaluates the overall effectiveness of \ourmodel{} in terms of both inference efficiency (speedup, tokens/s) and edit quality (Pass@1) compared to autoregressive decoding and state-of-the-art acceleration baselines across diverse code editing benchmarks.

% \textbf{RQ2:} How does each component of \ourmodel{} contribute to its performance? This question investigates the individual impact of \ourmodel{}'s core components—namely, \textit{Locate: Self-Editing Location with Code Reuse}, \textit{Edit-Oriented Fine-Tuning} for the draft model, and \textit{Entropy-Aware Dynamic Verification}—on its overall performance gains.

% \textbf{RQ3:} How effective is Entropy-Aware Dynamic Verification compared to other draft verification methods? This question compares our proposed verification mechanism against alternative strategies, such as greedy verification and top-k verification, in terms of balancing inference speed and output quality.

% \textbf{RQ4:} How generalizable is \ourmodel{} in code editing tasks with different code reuse rates? This question examines the performance of \ourmodel{} across scenarios with varying proportions of reused versus newly generated code, assessing its robustness and efficiency under different editing intensities.

\subsection{RQ1: Overall Performance of \ourmodel{}}

\begin{table*}[t]
    \centering
    \setlength{\abovecaptionskip}{0.1cm}
    \caption{The overall performance of \ourmodel{} and the baselines on code editing datasets.}
    \resizebox{\linewidth}{!}{ 
    \begin{tabular}{llcccccccccccc}
        \toprule
        \multirow{3}{*}{\textbf{Backbone}} & \multirow{3}{*}{\textbf{Approach}} & \multicolumn{6}{c}{\textbf{CanItEdit}}  & \multicolumn{6}{c}{\textbf{CodeIF-Bench}} \\
        \cmidrule(lr){3-8}  \cmidrule(lr){9-14}
        & &   \multicolumn{3}{c}{\textbf{Lazy}} & \multicolumn{3}{c}{\textbf{Descriptive}} &  \multicolumn{3}{c}{\textbf{L-1}} & \multicolumn{3}{c}{\textbf{L-2}} \\
        \cmidrule(lr){3-8} \cmidrule(lr){9-14}
         & &  {\textbf{Token/s}} & \textbf{Speedup} & \textbf{Pass@1} &  \textbf{Token/s} & \textbf{Speedup} & \textbf{Pass@1} &  \multicolumn{1}{c}{\textbf{Token/s}} & \textbf{Speedup} & \textbf{Pass@1} &  \textbf{Token/s} & \textbf{Speedup} & \textbf{Pass@1}\\
        \midrule
        \multirow{5}{*}{Qwen2.5-Coder}
        & AR & 13.0 & 1.00$\times$ & 54.3 & 13.0 & 1.00$\times$  & 68.6 & 13.4 & 1.00$\times$ & 43.1 & 12.7 & 1.00$\times$ &  4.2 \\
        % & SD &  21.7 & 1.67$\times$ & 54.3 & 22.0 & 1.69$\times$ & 68.6 & 22.8 & 1.70$\times$ & 43.1 & 22.5 & 1.77$\times$ & 4.2 \\
        & SD &  22.9 & 1.76$\times$ & 54.3 & 23.2 & 1.78$\times$ & 68.6 & 23.1 & 1.73$\times$ & 43.1 & 22.6 & 1.78$\times$ & 4.2 \\
        % & FastFixer & \underline{36.5} & 2.81$\times$ & 54.3 & \underline{37.1}  & 2.85$\times$ & 68.6 & \underline{59.1} & 4.41$\times$ & 43.1& \underline{97.5}& 7.68$\times$&4.2\\
        & FastFixer & \underline{37.1} & 2.85$\times$ & 54.3 & \underline{38.6}  & 2.97$\times$ & 68.6 & \underline{60.8} & 4.54$\times$ & 43.1& \underline{97.0}& 7.64$\times$&4.2\\
        & Ouroboros & - & - & - & - & - & - & - & - & - & - & - & -\\
         % & k=3 & \textbf{74.8} & \textbf{5.75$\times$} & 54.3 &  \textbf{72.4} & \textbf{5.57$\times$} & \textbf{69.5} & \textbf{87.8} & \textbf{6.55$\times$} & \textbf{43.3} & \textbf{113.7} & \textbf{8.95$\times$} & \textbf{7.2}\\
        % & k=3-trick1 & 67.3 &  & 55.2 &  67.6 &  & 70.5 \\
        & \ourmodel{} & \textbf{70.7}\ua{90.6\%} & \textbf{5.44$\times$} & \textbf{55.2} &  \textbf{70.0}\ua{81.3\%} & \textbf{5.38$\times$} & \textbf{70.5} & \textbf{81.4}\ua{33.9\%} & \textbf{6.07$\times$} & 43.1 & \textbf{104.9}\ua{8.1\%} & \textbf{8.26$\times$} & \textbf{4.5}\\
        % & k=5-trick1 & 71.7 &  & 53.3 &   72.3& & 68.6\\
        % & k=5-trick2 & 71.2 &  & 55.2 &  69.4 &  & 65.4 \\
         \midrule
        \multirow{5}{*}{DeepSeek-Coder} 
        & AR & 11.8 & 1.00$\times$ & 49.5 & 11.8 & 1.00$\times$ & 60.0 & 11.6 & 1.00$\times$ & 35.8 & 12.0 & 1.00$\times$ & 8.4   \\
        % & SD & 18.9 & 1.60$\times$ & 49.5 & 19.1 & 1.62$\times$ & 60.0 & 18.6 & 1.60$\times$ & 35.8 & 18.2 & 1.52$\times$ & 8.4  \\
        & SD & 19.6 & 1.67$\times$ & 49.5 & 19.9 & 1.69$\times$ & 60.0 & 18.5 & 1.59$\times$ & 35.8 & 18.7 & 1.56$\times$ & 8.4  \\
        % & FastFixer &56.6 & 4.80$\times$ & 49.5 & 51.7 & 4.38$\times$ & 59.0 & 40.5 & 3.49$\times$  & 35.8 & 74.6 & 6.21$\times$ & \\
        % & FastFixer & \underline{68.9} & 5.84$\times$ & 49.5 & \underline{59.8} & 5.07$\times$ & 60.0 & \underline{92.6} & 7.98$\times$  & 35.8 & \underline{99.1} & 8.26$\times$ & 8.4\\
        & FastFixer & \underline{77.9} & 6.60$\times$ & 49.5 & \underline{61.7} & 5.23$\times$ & 60.0 & \underline{93.7} & 8.08$\times$  & 35.8 & \underline{128.5} & 10.71$\times$ & 8.4\\
        & Ouroboros & 36.1 & 3.06$\times$  & 49.5 & 36.1 &3.06$\times$ & 60.0 & 31.8 & 2.74$\times$ & 35.8 & 33.2 & 2.77$\times$& 8.4 \\
         % & SpecFusion & \textbf{99.9} & \textbf{8.46}$\times$ & 48.6 & \textbf{73.2} & \textbf{6.85}$\times$ & 59.0 & 33.6 & 2.90$\times$ & 35.8 & - \\
        % & k=3 & 111.6 &  & 47.6 &  90.9 &  & 57.14 \\
        % & k=5 & 112.1 &  & 46.7 &  89.9 &  & 58.14 \\
        % & k=3-trick1 & 105.0 &  & 47.6 &  93.4 &  & 54.3 \\
        % & k=3-trick2 & 92.1 &  & 49.5 &  76.8 &  & 56.2 \\
        % & k=5-trick1 & 109.2 &  & 46.7 &  90.3 &  & 60.1 \\
        & \ourmodel{} & \textbf{122.5}\ua{57.3\%} & \textbf{10.38$\times$} & 48.5 &  \textbf{94.6}\ua{53.3\%} & \textbf{8.02$\times$} & 59.0 & \textbf{111.7}\ua{19.2\%} &\textbf{9.63$\times$} & \textbf{36.5} & \textbf{157.1}\ua{22.3\%} & \textbf{13.09$\times$}  & \textbf{8.7}\\

        \bottomrule
    \end{tabular}
    }
    \label{tab: main_results}
    \vspace{-0.3cm}
\end{table*}
To answer this research question, we compared our approach with three state-of-the-art inference acceleration baseline methods: SD, FastFixer, and Ouroboros. AR represents the autoregressive decoding results with KV-Cache of target models. Except for FastFixer, all other methods require the draft model. 
% Additionally, we include results from an autoregressive implementation with only KV-Cache optimization for reference. 
The results are presented in Table \ref{tab: main_results}. Due to implementation constraints in the original Ouroboros framework that preclude compatibility with Qwen-series models, we have not reported its results on the Qwen series models.

In terms of inference efficiency, our model surpasses all baselines, achieving highest inference acceleration of 8.26$\times$ and 13.09$\times$ on the Qwen2.5-Coder configuration and DeepSeek-Coder configuration. In addition, compared to the best baseline FastFixer, we can even improve the efficiency by up to 90.6\% in Qwen2.5-Coder configuration. Among all baselines, FastFixer delivers the best performance. We attribute this to the high similarity between program repair and code editing tasks, enabling its retrieval-based approach to also achieve gains in code editing tasks. Although this method achieves improved acceleration by reusing reusable code segments without relying on draft models, its performance degrades on data with autoregressive edit generation and excessive retrieval verification overhead. 
% This is because it depends on autoregressive patch generation by the target model and limits the number of token retrieval and reuse, thereby limiting the efficiency of code reuse. 
Ouroboros outperforms SD by reusing the verification process and historical context phrases to improve draft generation speed, achieving overall acceleration. However, it remains limited by the computational overhead of draft generation and repeatedly generated `redundant' code segments. In contrast, our approach dynamically and efficiently reuses reusable code from code to be edited, requiring only milliseconds of target model for a forward inference time. Furthermore, by optimizing the SD-based inference acceleration scheme at edit locations, our approach balances speed and effectiveness. As a result, \ourmodel{} demonstrates superior robustness and achieves the best acceleration performance across scenarios—whether the general function-level and class-level editing tasks CanItEdit or dialogue editing tasks CodeIF-Bench.
% which includes function-level and repo-level.

In terms of edit quality (functional correctness of the edited code), \ourmodel{} achieves an effective balance between acceleration and edit quality. We use the greedy decoding results of the target model as the quality baseline to verify whether our dynamic threshold validation affects the generation quality. In all model combinations, we maintained quality consistent with greedy decoding in most cases. This indicates that our dynamic threshold verification method can further improve speed without compromising generation quality. In addition, in some cases, it can even surpass the greedy decoding results. Specifically, Qwen2.5-Coder configuration even surpasses results of the target model using greedy decoding in quality (e.g. 55.2 vs 54.3 and 70.5 vs 68.6 on CanItEdit). Notably, the training data was annotated solely using the Qwen-Coder-32B-Instruct model without employing stronger closed-source models, yet still delivers improved results. While larger models typically exhibit superior code editing capabilities, we found that that draft models (especially the strong Qwen2.5-Coder) can also edit some data incorrectly edited by the target model alone. Our dynamic threshold validation effectively combines the strengths of both small and large models, achieving editing quality that sometimes surpasses the target model alone. This holds also for the DeepSeek-Coder configuration, which shows a marginal performance decrease CanItEdit because the inherent differences in editing capabilities between the target and draft model but matches greedy decoding overall and outperforms it on CodeIF-Bench. These results demonstrate our method's generalisation across model combinations and ability to maintain the acceleration-quality trade-off across diverse editing tasks.

\begin{table*}[t]
    \centering
    \setlength{\abovecaptionskip}{0.1cm}
    \caption{Ablation study results on code editing datasets.}
    \resizebox{\linewidth}{!}{ 
    \begin{tabular}{llcccccccccccc}
        \toprule
        \multirow{3}{*}{\textbf{Backbone}} & \multirow{3}{*}{\textbf{Approach}} & \multicolumn{6}{c}{\textbf{CanItEdit}}  & \multicolumn{6}{c}{\textbf{CodeIF-Bench}} \\
        \cmidrule(lr){3-8}  \cmidrule(lr){9-14}
        & &   \multicolumn{3}{c}{\textbf{Lazy}} & \multicolumn{3}{c}{\textbf{Descriptive}} &  \multicolumn{3}{c}{\textbf{L-1}} & \multicolumn{3}{c}{\textbf{L-2}} \\
        \cmidrule(lr){3-8} \cmidrule(lr){9-14}
         & &  {\textbf{Token/s}} & \textbf{Speedup} & \textbf{Pass@1} &  \textbf{Token/s} & \textbf{Speedup} & \textbf{Pass@1} &  \multicolumn{1}{c}{\textbf{Token/s}} & \textbf{Speedup} & \textbf{Pass@1} &  \textbf{Token/s} & \textbf{Speedup} & \textbf{Pass@1}\\
        \midrule
        \multirow{5}{*}{Qwen2.5-Coder}
        % & AR & 13.0 & 1.00$\times$ & 54.3 & 13.0 & 1.00$\times$  & 68.6 & 13.4 & 1.00$\times$ & 43.1 & 12.7 & 1.00$\times$ &  3.5 \\
        & Base+Greedy &  21.7 & 1.67$\times$ & 54.3 & 22.0 & 1.69$\times$ & 68.6 & 22.8 & 1.70$\times$ & 43.1 & 22.5 & 1.77$\times$ & 4.2 \\
        & Reuse+Base+Greedy & 60.5 & 4.65$\times$ & 54.3 & 63.5 & 4.88$\times$ & 68.6 & 70.6 & 5.27$\times$ & 43.1 & 88.0 & 6.93$\times$ & 4.2\\

        & Reuse+Our+Greedy & 63.7 & 4.90$\times$ & 54.3 &  65.9 & 5.07$\times$ & 68.6 & 79.9 & 5.96$\times$ & 43.1 & 99.8 & 7.86$\times$ & 4.2 \\
        & Reuse+SFT+Entropy &  70.4 & 5.42$\times$ & \textbf{56.2} & 68.3 & 5.25$\times$ & \textbf{70.5} & 75.7 &5.65$\times$ & 42.9 & 100.3 & 7.90$\times$ & 4.2 \\
        % & Editing Reuse+Base+Entropy & 67.5 &  67.5 & 54.3 & 65.8 & - & 68.6 & - & - & - & - & - & -\\
        % & Reuse+SFT+Entropy & 70.5 & 5.42$\times$ & \textbf{56.1} &  68.3 & 5.25$\times$ & \textbf{70.5} & \textbf{81.4} & \textbf{6.07$\times$} & 43.1 & \textbf{104.9} & \textbf{8.26$\times$} & \textbf{4.5}\\
        
        & Reuse+Our+Entropy & \textbf{70.7} & \textbf{5.44$\times$} & 55.2 &  \textbf{70.0} & \textbf{5.38$\times$} & \textbf{70.5} & \textbf{81.4} & \textbf{6.07$\times$} & 43.1 & \textbf{104.9} & \textbf{8.26$\times$} & \textbf{4.5}\\

         \midrule
        \multirow{4}{*}{DeepSeek-Coder} 
        % & AR & 11.8 & 1.00$\times$ & 49.5 & 11.8 & 1.00$\times$ & 60.0 & 11.6 & 1.00$\times$ & 35.8 & 12.0 & 1.00$\times$ & 5.3   \\
        & Base+Greedy & 18.9 & 1.60$\times$ & 49.5 & 19.1 & 1.62$\times$ & 60.0 & 18.6 & 1.60$\times$ & 35.8 & 18.2 & 1.52$\times$ & 8.4  \\
 
        & Reuse+Base+Greedy & 97.8 & 8.29$\times$ & 49.5 & 79.3 & 6.72$\times$ & 60.0 & 103.9 & 8.96$\times$ & 35.8 & 138.1 & 11.51$\times$ & 8.4 \\

        & Reuse+Our+Greedy &  105.1 & 8.91$\times$ & 49.5 & 85.0 & 7.20$\times$ & 60.0 & 105.9 & 9.13$\times$ & 35.8 & 140.2 & 11.68$\times$& 8.4\\

        % & Editing Reuse+Base+Entropy & 130.7 & 11.08$\times$ & 47.6 &  93.5 & 7.92$\times$ & 53.3 & 113.7 & 9.80$\times$ & - & 156.7 & 13.06$\times$ & -\\
        
        % & Reuse+SFT+Entropy & \textbf{127.1} & \textbf{10.77$\times$} & 43.8 &  \textbf{90.7} & \textbf{7.68$\times$} & 54.3 & \textbf{111.7} &\textbf{9.63$\times$} & \textbf{36.5} & \textbf{157.1} & \textbf{13.09$\times$}  & \textbf{8.7}\\
        & Reuse+SFT+Entropy & \textbf{127.1} & \textbf{10.77$\times$} & 43.8 & 90.7 & 7.69$\times$ & 54.3 & \textbf{112.6} & \textbf{9.71$\times$} & \textbf{37.0} & \textbf{157.7} & \textbf{13.14$\times$} & \textbf{9.8}  \\
        & Reuse+Our+Entropy & 122.5 & 10.38$\times$ & 48.5 &  \textbf{94.6} & \textbf{8.02$\times$} & 59.0 & 111.7 &9.63$\times$ & 36.5 & 157.1 & 13.09$\times$  & 8.7\\

        \bottomrule
    \end{tabular}
    }
    \label{tab: ablation_study}
    % \vspace{-0.2cm}
\end{table*}
\begin{RQbox} \textbf{RQ1 Summary: }
Our method substantially outperforms existing SOTA acceleration methods. Compared to the best-performing baseline FastFixer, \ourmodel{} improves efficiency by an average of \textbf{70.6\%} on CanItEdit and \textbf{21.6\%} on CodeIF-Bench. Concurrently, \ourmodel{} preserves—and in many cases even surpasses—the generation quality of greedy decoding.
\end{RQbox}

\subsection{RQ2: Contributions of Each Component}

\begin{table*}[t]
    \centering
    \setlength{\abovecaptionskip}{0.1cm}
    \caption{Comparison results of different verification methods. The model combination is DeepSeek-Coder.}
    \resizebox{\linewidth}{!}{ 
    \begin{tabular}{lcccccccccccc}
        \toprule
          \multirow{3}{*}{\textbf{Approach}} & \multicolumn{6}{c}{\textbf{CanItEdit}}  & \multicolumn{6}{c}{\textbf{CodeIF-Bench}} \\
        \cmidrule(lr){2-7}  \cmidrule(lr){8-13}
         &   \multicolumn{3}{c}{\textbf{Lazy}} & \multicolumn{3}{c}{\textbf{Descriptive}} &  \multicolumn{3}{c}{\textbf{L-1}} & \multicolumn{3}{c}{\textbf{L-2}} \\
        \cmidrule(lr){2-7} \cmidrule(lr){8-13}
          &  {\textbf{Token/s}} & \textbf{Speedup} & \textbf{Pass@1} &  \textbf{Token/s} & \textbf{Speedup} & \textbf{Pass@1} &  \multicolumn{1}{c}{\textbf{Token/s}} & \textbf{Speedup} & \textbf{Pass@1} &  \textbf{Token/s} & \textbf{Speedup} & \textbf{Pass@1}\\
        \midrule
         Greedy &  105.1 & 8.91$\times$ & \textbf{49.5} & 85.0 & 7.20$\times$ & \textbf{60.0} & 105.9 & 9.13$\times$ & 35.8 & 140.2 & 13.32$\times$& 8.4\\
        Top-k=3 & 128.1 & 10.86$\times$ & 44.8 & 102.5 & 8.68$\times$ & 50.5 & 112.9&9.78$\times$ & 35.6 & 159.8 & 13.31$\times$  & 9.3\\
        % & FastFixer &56.6 & 4.80$\times$ & 49.5 & 51.7 & 4.38$\times$ & 59.0 & 40.5 & 3.49$\times$  & 35.8 & 74.6 & 6.21$\times$ & \\
        Top-k=5 & 131.4 & 11.14$\times$ & 43.8 & \textbf{102.7} & \textbf{8.70$\times$} & 48.6 &  113.4 & 
        9.77$\times$ & 35.4 & 160.8 & 13.43$\times$ &  \textbf{9.8}
        \\
        Direct &  \textbf{131.9} & \textbf{11.18$\times$} &43.8 &102.0 & 8.64$\times$ & 47.6 & \textbf{114.4} & \textbf{9.86$\times$} & 35.1 & \textbf{161.2} & \textbf{13.89$\times$} & \textbf{9.8} \\
        Entropy & 122.5 & 10.38$\times$ & 48.5 &  94.6 & 8.02$\times$ & 59.0 & 111.7 &9.63$\times$ & \textbf{36.5} & 157.1 & 13.09$\times$  & 8.7\\

        \bottomrule
    \end{tabular}
    }
    \label{tab: compare_decoding}
    \vspace{-0.3cm}
\end{table*}
To assess the contribution of each component, we perform ablation studies on \ourmodel{}. Specifically, we adopt the basic SD method as the baseline and incrementally incorporate our components, analyzing the performance gains at each stage. The experimental results are presented in \Cref{tab: ablation_study}. For clarity, we decompose SD into two key aspects: draft model type and decoding method. Here, \textit{Base} denotes the original base model serving as the draft model, \textit{Greedy} indicates greedy decoding verification, \textit{Reuse} represents the module that reuses code from the code to be edited, \textit{Our} refers to the draft model fine-tuned with our edit-oriented loss (\Cref{eq:masked-loss}), \textit{SFT} refers to a draft model fine-tuned with standard supervised fine-tuning loss, and \textit{Entropy} signifies our Entropy-Aware Dynamic Verification. 

The experimental results demonstrate that each module of \ourmodel{} contributes to performance improvement. Notably, the most substantial gains occur after integrating the \textit{Reuse} module. This module reuses a large number of tokens from the original code through a single forward inference of the target model, thereby greatly reducing generation time. Additionally, it lightly locates potential editing positions and automatically controls whether to reuse code or efficiently generate edited content, highlighting the role of the \textit{Reuse} module as a key component of \ourmodel{}. 

We present comparison results with standard fine-tuning methods (\textit{SFT}) versus our edit-oriented fine-tuning (\textit{Our}). Except for the training loss, all other experimental settings remain identical. The results demonstrate that traditional fine-tuning methods exhibit lower robustness. Specifically, in the Qwen2.5-Coder model ensemble, \textit{SFT} achieves comparable performance to \textit{Our} on the CanItEdit benchmark but underperforms on CodeIF-Bench. Conversely, in the DeepSeek-Coder model ensemble, traditional fine-tuning leads to significantly degraded generation quality on CanItEdit while outperforming \ourmodel{} on CodeIF-Bench. We hypothesize that the loss functions in traditional fine-tuning lack task-specific optimization, causing the model to potentially learn to exploit redundant code generation to minimize training loss, thereby limiting generalization across benchmarks. In contrast, \ourmodel{}'s targeted loss training enhances performance at edit patch positions and improves overall effectiveness.

We further observe that while the fine-tuned draft model enhances acceleration, the improvements are modest if only fine-tuning is applied without other optimizations. This suggests that parameter discrepancies between the draft and target models can lead to output distribution differences that may not be fully resolved through fine-tuning alone. However, the \textit{Entropy} verification method provides additional acceleration benefits and generally surpasses greedy decoding in performance. These results support our hypothesis: strict alignment between the draft model's output and the target model's distribution is not essential for maintaining quality if a smart verification strategy is used. The \textit{Entropy} method effectively optimizes the trade-off between generation quality and acceleration, leading to performance enhancement.
\begin{RQbox} \textbf{RQ2 Summary: }
The \textit{Reuse} module represents the most crucial component of \ourmodel{} for speedup. \textit{Edit-Oriented Fine-Tuning} and \textit{Entropy-Aware Dynamic Verification} further enhance \ourmodel{}'s overall performance by accelerating edit generation.
\end{RQbox}
\subsection{RQ3: Comparison with Different Draft Verification Methods}
To compare different draft verification methods within the \textit{Generation} phase of \ourmodel{}, we evaluate the following approaches:
\begin{itemize}
    \item Greedy: The draft model's greedy decoding output must exactly match the target model's greedy decoding output for acceptance.
    \item Direct: The draft model's greedy decoding outputs are used directly without any verification by the target model.
    \item Top-k: The draft model's greedy decoding output is accepted if it appears among the top-k tokens (ranked by probability) in the target model's output distribution. We test with a fixed $k=3$ and $k=5$.
    \item Entropy (Ours): Our Entropy-Aware Dynamic Verification.
\end{itemize}

As shown in Table \ref{tab: compare_decoding}, when the verification threshold decreases progressively from \textit{Greedy} to \textit{Direct}, the draft model's acceleration effect on the target model improves, but at the cost of reduced generation quality. When the validation threshold is lowered, the target model accepts more edited content from the draft model, thereby improving generation speed. Because validation failure not only decreases the number of valid tokens but also forces the draft model to regenerate draft, impairing efficiency. However, due to the inherent differences in editing capabilities between the target model and the draft model, this efficiency gain comes at the expense of significantly degraded generation quality, evidenced by Pass@1 drops of approximately 6\% and 13\% on the CanItEdit. These findings reinforce our motivation: balancing acceleration and quality. 

Our proposed validation method achieves this balance, delivering superior acceleration compared to the greedy algorithm while maintaining—and in some cases surpassing—its generation quality. We attribute this success to dynamic thresholding, which improves the draft acceptance rate, and effectively combines the editing strengths of both target and draft models to enhance quality. Notably, on the CodeIF-Bench, L-2 generation quality improves as the threshold decreases. We hypothesize that this results from the draft model's strong L-2 editing capabilities, further refined through supervised fine-tuning. Our validation method also optimally leverages this characteristic, maintaining generation quality between the draft and target models, thus further demonstrating its effectiveness.

\begin{RQbox} \textbf{RQ3 Summary: } 
Our \textit{Entropy-Aware Dynamic Verification} method not only effectively improves the draft model's acceleration effect on the target model, but also maintains - and in some cases surpasses - the generation quality of greedy decoding validation. 

% Our \textit{Entropy-Aware Dynamic Verification} method effectively balances the draft model's acceleration effect on the target model and the generation quality across all benchmarks, often outperforming fixed Top-k strategies and, in some cases, even surpassing the quality of strict greedy decoding validation while providing better speedup.\vspace{-0.3cm}
\end{RQbox}

\subsection{RQ4: Performance with Different Code Reuse Rates}

\begin{figure}[!ht]
    \centering
    \begin{subfigure}[b]{0.45\textwidth}
        \includegraphics[width=\textwidth]{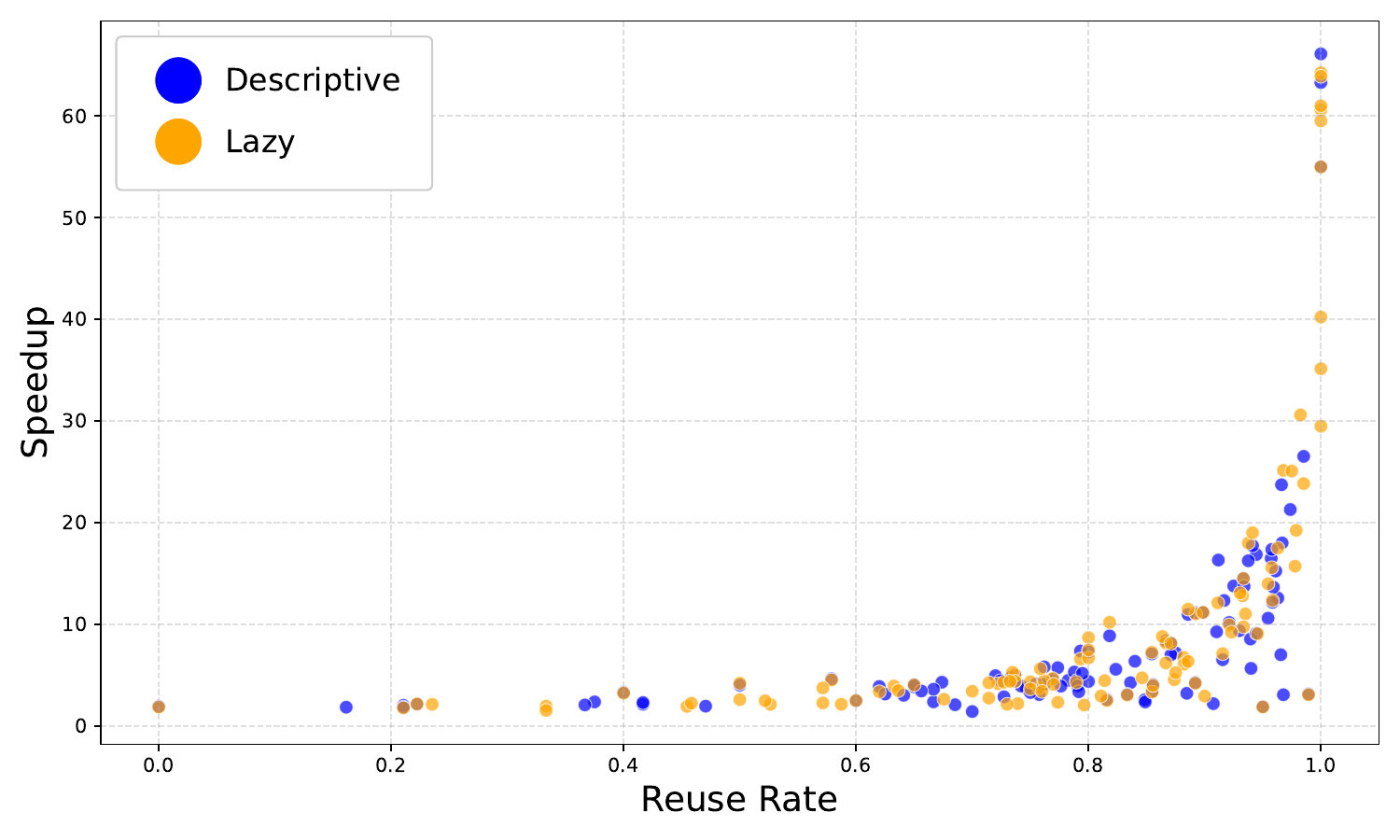}
        \caption{DeepSeek-Coder}
        \label{fig:canitedit_deep}
    \end{subfigure}
    \hfill
    \begin{subfigure}[b]{0.45\textwidth}
        \includegraphics[width=\textwidth]{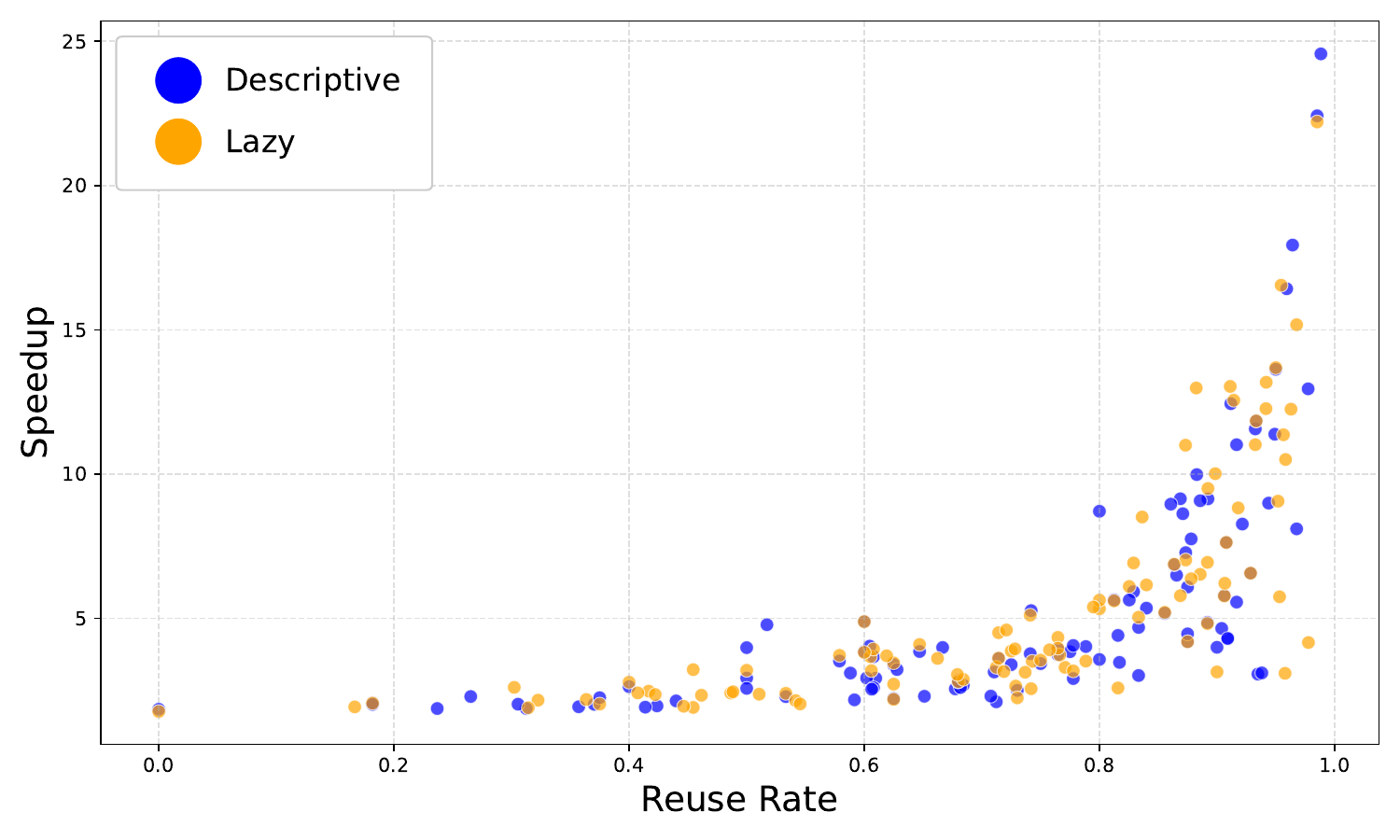}
        \caption{Qwen2.5-Coder}
        \setlength{\belowcaptionskip}{-0.15cm} 
        \label{fig:canitedit_qwen}
    \end{subfigure}
    % \vspace{-1mm}
    % \setlength{\abovecaptionskip}{0.05cm}  
    % \setlength{\belowcaptionskip}{-0.05cm} 
    \caption{Inference speedup of \ourmodel{} with different reuse rates on CanItEdit. The reuse rate is the percentage of tokens in the final edited code that come from the original code.}
    \label{fig: code_reuse_rate}
    % \vspace{-0.3cm}
\end{figure}

We further investigate the impact of code reuse rates on \ourmodel{}'s performance. As shown in \Cref{fig: code_reuse_rate}, higher code reuse rates generally correlate with increased acceleration ratios. This is because reusable code is often block-based, and \ourmodel{}'s \textit{Reuse} phase can efficiently reuse a large number of tokens in a single target model forward pass, thereby greatly improving efficiency. However, we observe that in a few cases, substantial increases in reuse rates lead to only marginal improvements in acceleration. We attribute this to the fragmentation of reusable code in these instances, which can result in only a small amount of code tokens being reused at a time and increase the overhead of switching between \textit{Reuse} and \textit{Generate}. Meanwhile, even at low reuse rates (e.g., \~0.5), our method achieves measurable acceleration, demonstrating robustness.

Performance varies significantly across different model families. The DeepSeek-Coder combination achieves peak acceleration exceeding 50$\times$ at a 100\% reuse rate, while Qwen2.5-Coder's maximum acceleration remains below 25$\times$. This disparity results in DeepSeek-Coder's average acceleration ratio appearing higher than Qwen2.5-Coder's at very high reuse rates. Our analysis suggests this anomaly (extremely high speedup at 100\% reuse for DeepSeek-Coder) occurs partly because DeepSeek-Coder, being inherently somewhat weaker in instruction following for certain complex edits than Qwen2.5-Coder in our experiments, occasionally tends to ignore editing instructions and output the unmodified code. In such 100\% reuse scenarios, \ourmodel{} very efficiently verifies the entire original code as correct with minimal overhead. Notably, when changing instructions from ``lazy'' to ``descriptive'' formats in CanItEdit, the frequency of 100\% reuse cases for DeepSeek-Coder decreases significantly, indicating its greater sensitivity to instruction specificity. In contrast, Qwen2.5-Coder demonstrates superior robustness in adhering to edit instructions, rarely exhibiting 100\% code reuse regardless of instruction style, leading to more consistent speedup patterns.

\begin{RQbox} \textbf{RQ4 Summary: }
In most scenarios, \ourmodel{} demonstrates a positive correlation between code reuse rate and acceleration ratio, while still demonstrates good acceleration in lower reuse rate. Model-specific behaviors can influence performance at extreme reuse rates.
\end{RQbox}

% #with particularly significant improvements observed when reused code appears in contiguous blocks.
\section{Discussion}

\subsection{Case Study}

\begin{figure}[!ht]
\centering
\setlength{\abovecaptionskip}{0.2cm}
\includegraphics[width=\linewidth]{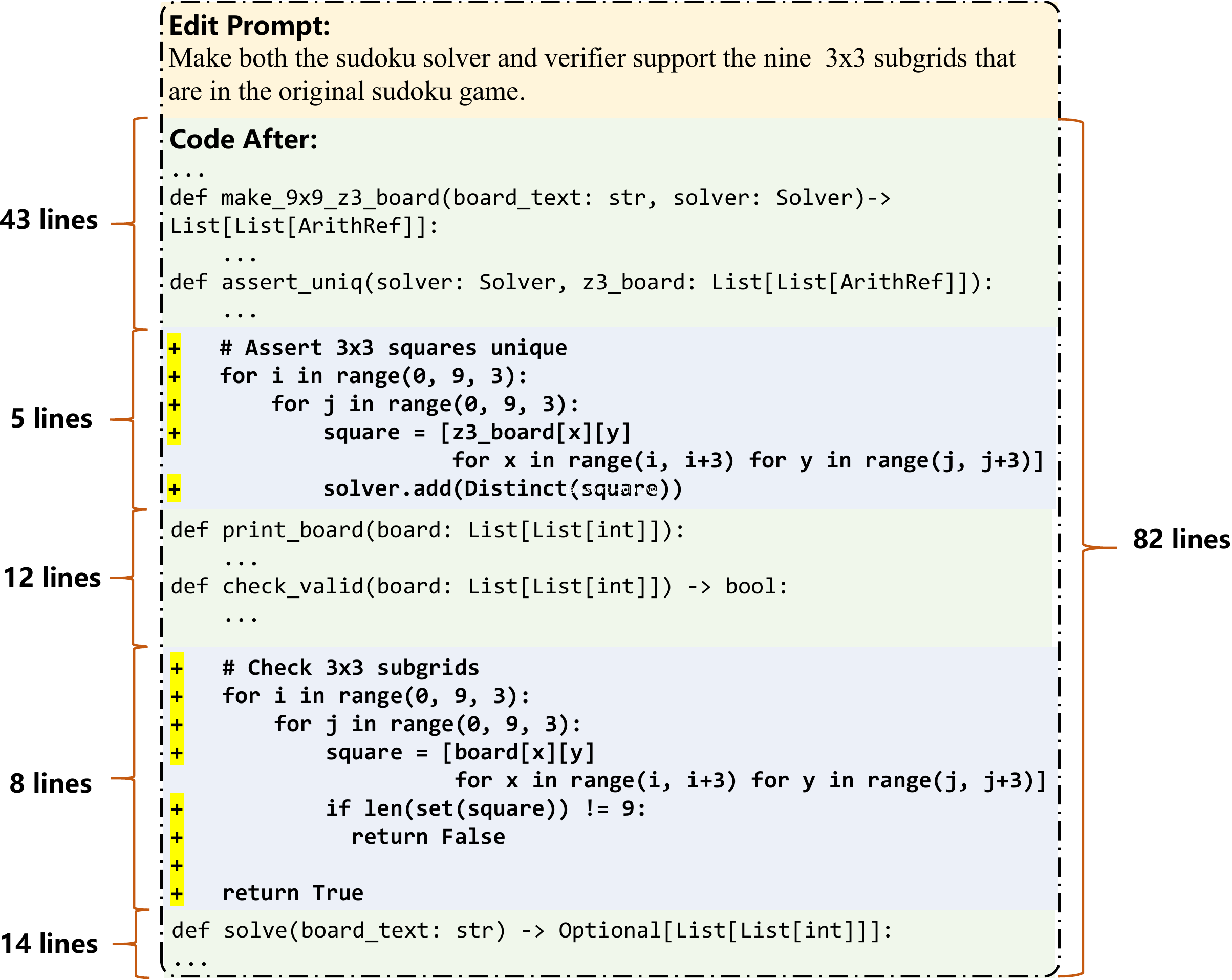}
\caption{Case study. An example generated by the Qwen2.5-Coder configuration from CanItEdit (task ID 25). The proportion of newly edited lines (patches) is approximately 16\% of the final code.}
\label{fig:case_study}
\vspace{-0.3cm}
\end{figure}

We demonstrate the effectiveness of our method through the case study presented in \Cref{fig:case_study}. The task involves implementing functional enhancements to existing code, representing a challenging code editing scenario where: (1) the precise editing locations cannot be determined solely from the instruction, and (2) multiple modification points are required. When applying \ourmodel{}, the Qwen2.5-Coder configuration achieves a 5.2$\times$ speedup compared to traditional autoregressive methods, outperforming the best baseline (FastFixer at 3.3$\times$) up to 50\%. Qwen2.5-Coder-32B-Instructi requires 58.4 seconds for autoregressive decoding (using 4 RTX 4090 GPUs), while our method completes the same task in just 11.2 seconds, greatly improving editing efficiency. Our method provides superior performance through its two-phase design: (1) the \textit{Reuse} phase reuses large amounts of redundant code in a lightweight manner while simultaneously identifying potential editing locations, and (2) the \textit{Generate} phase enables efficient edit generation through high-quality draft models and an optimized verification algorithm that balances generation quality and efficiency. 

As illustrated in \Cref{fig:case_study}, \ourmodel{} achieves reuse of 69 lines of code through just 3 target model forward inferences in its \textit{Reuse} phase, reusing 43 lines, 12 lines, and 14 lines, respectively. The first two reuse inferences also effectively pinpoint edit locations where new code needs to be inserted. Subsequently, the combination of a high-quality draft model (fine-tuned for edits) and our dynamic threshold verification mechanism enables efficient generation of the new code segments.

\subsection{Hyperparameter Discussion}
\begin{figure}[!ht]
    \centering
    \begin{subfigure}[b]{0.49\textwidth}
        \includegraphics[width=\textwidth]{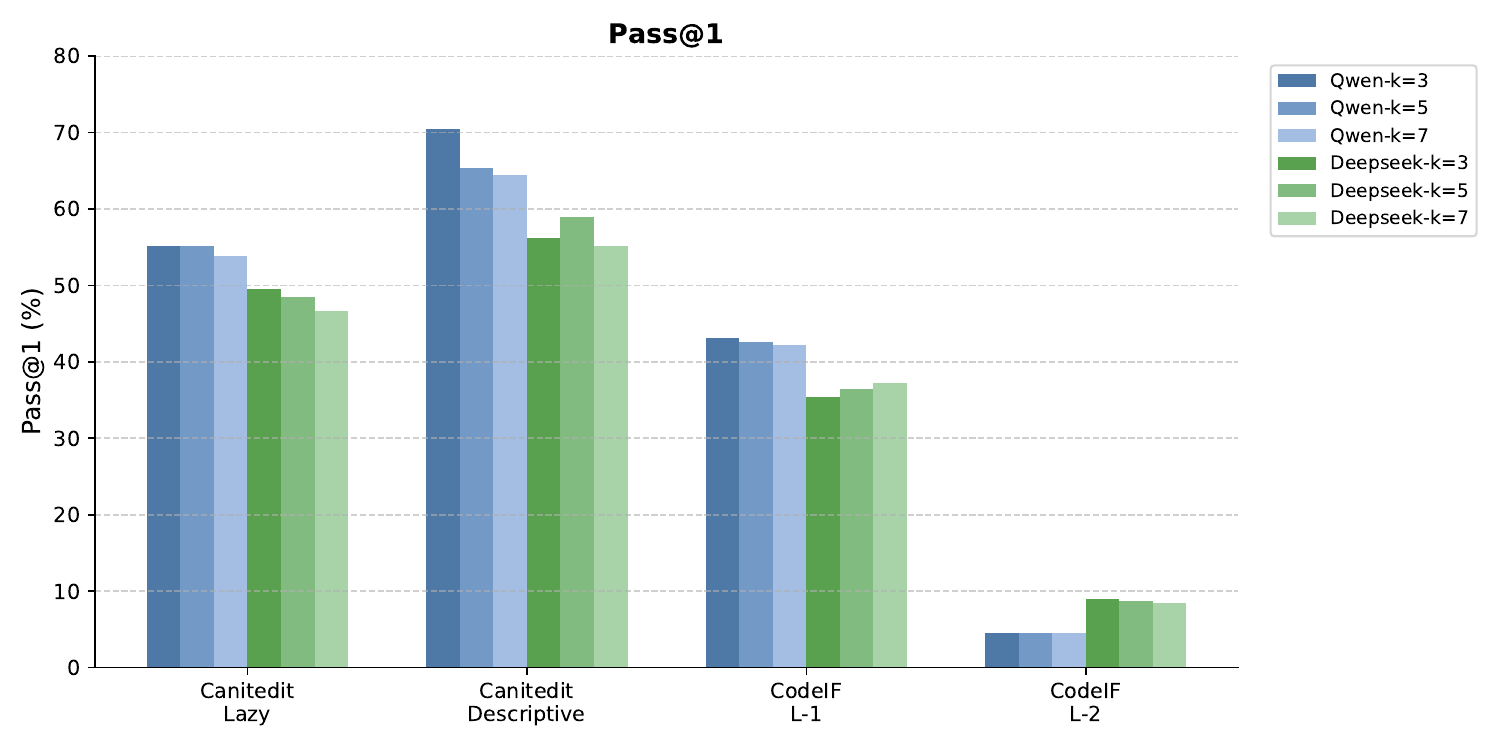} % Assuming this is Pass@1 for DeepSeek
        \caption{Pass@1}
        \label{fig:par_pass}
    \end{subfigure}
    \hfill
    \begin{subfigure}[b]{0.49\textwidth}
        \includegraphics[width=\textwidth]{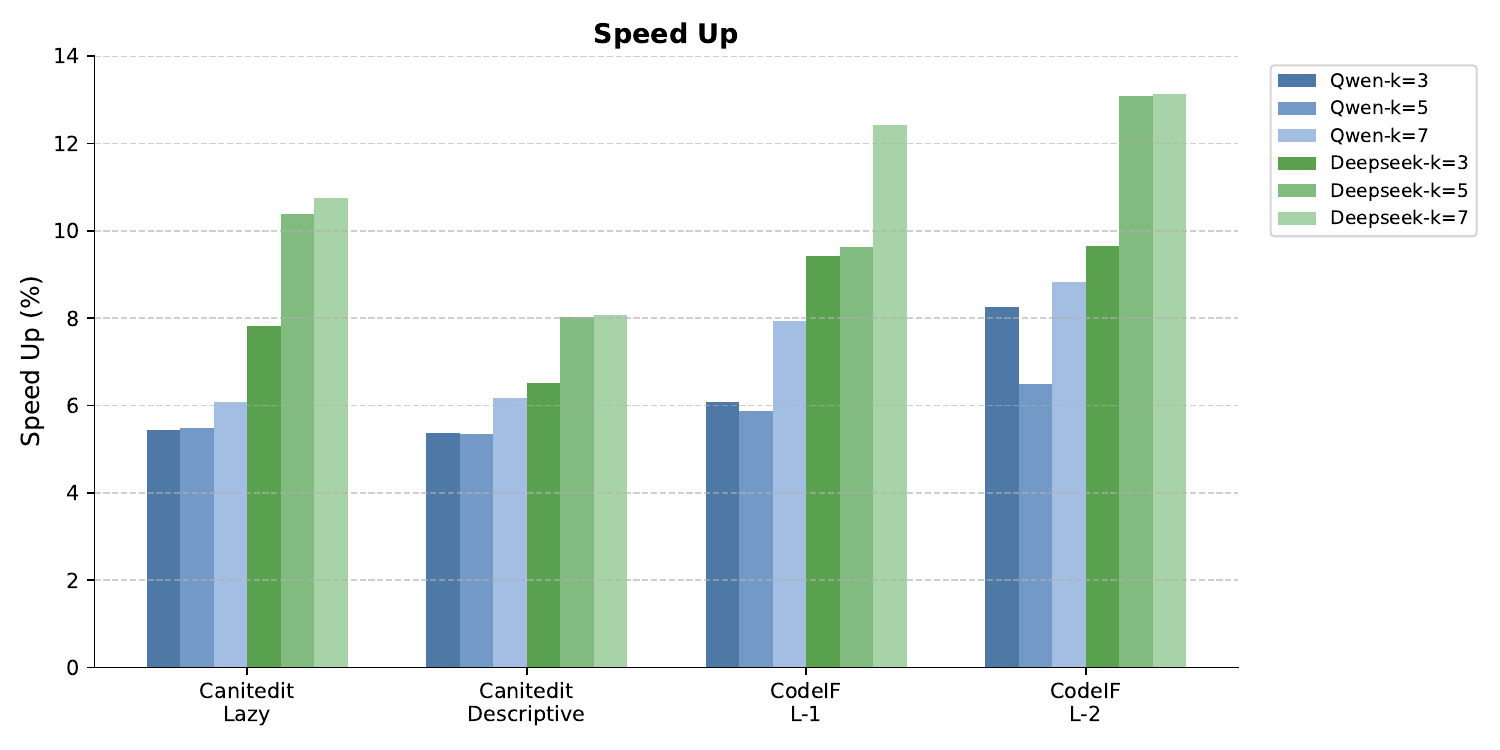}
        \caption{Speedup}
        \setlength{\belowcaptionskip}{-0.15cm} 
        \label{fig:par_speed}
    \end{subfigure}
    % \vspace{-1mm}
    % \setlength{\abovecaptionskip}{0.05cm}  
    % \setlength{\belowcaptionskip}{-0.05cm} 
    \caption{\ourmodel{} performance (Pass@1 and Speedup) at different base entropy thresholds $k$.}
    \label{fig: par}
    \vspace{-0.3cm}
\end{figure}

In \ourmodel{}, the base threshold $k$ plays a crucial role in the validation process. Figure \ref{fig: par} illustrates the performance of \ourmodel{} with different threshold values ($k=3,5,7$). The results reveal that: (1) the Qwen2.5-Coder configuration achieves optimal performance at $k=3$, while the DeepSeek combination performs best at $k=5$; (2) the acceleration ratio shows significant improvement with increasing threshold values; and (3) Pass@1 remains relatively stable across these tested thresholds, exhibiting only marginal variations. This indicates a degree of robustness in hyperparameter selection concerning generation quality within this range, allowing for tuning towards higher speedup without drastic quality drops. Our chosen values ($k=3$ for Qwen, $k=5$ for DeepSeek) represent a good balance found empirically.

\subsection{Comparison with Code Editing Methods}
Growing popularity of code editing techniques\cite{liu2024coedpilot,fastapply}, it is crucial to evaluate the efficiency of \ourmodel{} in comparison with existing approaches. Specifically, we select the following two representative methods: the open-source code editing tool FastApply\cite{fastapply} and the pretraining-based approach CoEdPilot\cite{liu2024coedpilot}. 

\subsubsection{Compared to CoEdPilot} CoEdPilot is built on encoder-decoder architecture like 
% CodeBERT\cite{feng2020codebert} and 
CodeT5\cite{wang2021codet5}, whereas EfficientEdit is designed for most LLMs based on decoder-only architecture (\textit{e.g.}, Qwen, DeepSeek, LLaMA). This architectural mismatch makes direct effectiveness comparison challenging, and retraining CoEdPilot on larger decoder-only models is beyond our current scope. Thus, in Table \ref{tab:compare_CoEdPilot}, we compare editing time under the same experimental setting, acknowledging that CoEdPilot (about 1.xB parameters) and EfficientEdit (about 40B total) differ greatly in model-scale. 

\begin{table}[h]
\centering\small
\setlength{\abovecaptionskip}{0.1cm}
\caption{Editing time ($s \downarrow$) of CoEdPilot and \ourmodel{} on CanItEdit dataset. The backbone is Qwen2.5-Coder combination.}
\begin{tabular}{lcc}
\toprule
 & \textbf{Lazy} &  \textbf{Descriptive} \\
  \cmidrule(lr){1-3}
 CoEdPilot   & 40.2  &   38.2    \\
 \ourmodel{}   & \textbf{7.0}  &   \textbf{7.3}    \\
\bottomrule
\end{tabular}
\label{tab:compare_CoEdPilot}
\end{table}
As seen from the results, EfficientEdit shows subtantially higher efficiency. Specifically, CoEdPilot uses a locator combined with previous location and editing operations to locate positions and uses a locator-generator loop with iterative forward passes, which introduces cumulative context and latency. Despite this, its localization-editing paradigm is promising. Since it still relies on autoregressive decoding, EfficientEdit or speculative decoding techniques could be integrated to further accelerate its pipeline.
\begin{table}[h]
\centering
\setlength{\tabcolsep}{10pt} % 调整列间距
\setlength{\abovecaptionskip}{0.2cm}
\caption{Editing efficiency comparison of FastApply and \ourmodel{} on the CanItEdit dataset. The backbone is Qwen2.5-Coder combination.}
\label{tab:compare_FastApply}
\resizebox{\linewidth}{!}{
\begin{tabular}{lcccc}
\toprule
& \multicolumn{2}{c}{\textbf{Lazy}} & \multicolumn{2}{c}{\textbf{Descriptive}} \\
\cmidrule(lr){2-3} \cmidrule(lr){4-5}
& Token/s & Speedup & Token/s & Speedup \\
\midrule
FastApply   & 18.9 & 1.45$\times$ & 17.7 & 1.36$\times$ \\
\ourmodel{} & \textbf{70.7} & \textbf{5.44$\times$} & \textbf{70.0} & \textbf{5.38$\times$} \\
\bottomrule
\end{tabular}
}
\end{table}
\subsubsection{Compared to FastApply} FastApply takes the original and diff-formatted code from a target-LLM as input and outputs full edited code. It achieves high throughput via commercial backends (\textit{e.g.}, SoftGen\cite{SoftGen}) that may optimize quantization, caching, or memory. To ensure fairness, we used the same environment as in our setting: Qwen2.5-Coder-32B-Instruct generated diff patches, and FastApply (7B), without any commercial tools, produced final outputs. Results are summarized as Table \ref{tab:compare_FastApply}. Under identical conditions, EfficientEdit achieves substantially higher token throughput and speedup than FastApply. While FastApply reduces redundancy via diff-based editing, it remains bottlenecked by autoregressive decoding, which generates one token per inference step. In contrast, EfficientEdit can verify and accept multiple tokens per step, breaking this limit and providing superior efficiency.

\section{Related Work}

\subsection{LLM for Code Editing}
% Recent advances in LLM-based code editing have primarily focused on end-to-end training paradigms. Models like StarCoder\cite{lozhkov2024starcoder}, OctoCoder\cite{muennighoff2024octopack}, and EditCoder\cite{canitedit} leverage commit histories during pre-training or fine-tuning to enhance editing capabilities. The development of InstructCoder\cite{instructcoder}, specifically designed for instructional code editing tasks, represents a notable advancement in this field. Subsequent improvements include NextCoder \cite{aggarwal2025nextcoder} which introduced an adaptive fine-tuning algorithm, and CodeEditor\cite{li2023codeeditor} which employed mutated code snippets in pre-training. Alternative approaches like CoEdPilot [7] utilize separate localization and generation models, though their autoregressive nature limits efficiency.

% These developments have been evaluated through comprehensive benchmarks including CanItEdit [3], EditEval [4], CodeEditorBench [8], SWE-Bench [9], and CodeIF-Bench [10], demonstrating substantial progress in the field. However, current approaches remain constrained by autoregressive decoding limitations and redundant code generation. In contrast, our \ourmodel{} method addresses these efficiency challenges while maintaining editing quality.

Recent benchmarks (CanItEdit \cite{canitedit}, EditEval \cite{instructcoder}, CodeEditorBench \cite{guo2024codeeditorbench}, SWE-Bench \cite{jimenez2024swe},CodeIF-Bench\cite{CodeIF-Bench}) demonstrate substantial progress in LLM-based code editing\cite{qwencoder,zhu2024deepseek,lozhkov2024starcoder}. Current approaches typically employ end-to-end training and inference paradigms, with models like StarCoder \cite{lozhkov2024starcoder}, OctoCoder \cite{muennighoff2024octopack}, and EditCoder \cite{canitedit} leveraging commit histories in pre-training or fine-tuning to enhance editing capabilities. The recent development of InstructCoder\cite{instructcoder}, a model specifically designed and evaluated for instructional code editing tasks, represents a notable advancement in LLM-based code modification research. NextCoder\cite{aggarwal2025nextcoder} enhance its end-to-end code editing performance by incorporating a novel adaptive fine-tuning algorithm during training. However, their efficiency remains constrained by autoregressive decoding limitations and generating redundant code. CodeEditor\cite{li2023codeeditor} enhances LLM-based code editing performance via additional pre-training on mutated code snippets.
Some solutions like CoEdPilot \cite{liu2024coedpilot} jointly train a localization LLM and a generation LLM, where the localization model pinpoints edit positions to improve the generation model's efficiency, but the autoregressive inference process in both models still constrains their overall editing efficiency. In contrast, we proposes the \ourmodel{} method to improve the inference efficiency of autoregressive LLMs in code editing. Combining \ourmodel{} maybe achieve more efficient and high-quality code editing.

\subsection{Efficient Inference for LLMs}

Most current LLMs rely on autoregressive decoding for inference, suffering from inefficient sequential token generation. Speculative decoding\cite{leviathan2023fastinferencetransformersspeculative} has emerged as a promising solution, significantly improving inference speed while maintaining output quality. In general-purpose generation, approaches like lookahead decoding\cite{lookahead} employed phrases to accelerate the target model, while Ouroboros\cite{zhao2024ouroboros} uses rejected drafts and historical content to improve the efficiency of draft model generating drafts. Judge Decoding\cite{bachmann2025judgedecodingfasterspeculative} enables the target model to recover correct but invalidated draft content via a trained discriminative module.
For code-related tasks, REST\cite{rest} leverages the Stack dataset as a retrieval repository to boost generation efficiency, while FastFixer\cite{fastfix} retrieves reusable fragments from historical error codes as drafts to enhance program repair efficiency. Building upon these advances in speculative decoding and leveraging the unique characteristics of code editing tasks, we design a dedicated inference acceleration method specifically optimized for code editing scenarios. 

\section{Threats to Validity}

% \textbf{External Validity} relate to the generalization capability of \ourmodel{}. We employed a combination of state-of-the-art backbone LLMs (Qwen and DeepSeek) to evaluate our method across diverse editing scenarios. Our evaluation encompasses function, class, and repository-level editing tasks using both the crowd-sourced CanItEdit dataset and the interactive CodeIF-Bench benchmark. This comprehensive evaluation significantly mitigates threats to external validity. 
\textbf{External Validity} relates to the generalization and overhead of \ourmodel{}. To assess generalization, we combine state-of-the-art backbones (Qwen and DeepSeek) and evaluate across function-, class-, and repository-level editing tasks on both CanItEdit and CodeIF-Bench. The results demonstrate \ourmodel{}'s excellent generalizability, thereby reducing threats to external validity. Regarding overhead, \ourmodel{} incurs two additional costs: (1) extra GPU memory to load the draft model during inference, and (2) training cost for the draft model. Yet, the draft model uses only $ \sim 22\%$ of the parameters and memory of the 32B target model, and training converges within 2 hours on two NVIDIA 4090 GPUs with training data easily generated by the target LLM. Despite this modest cost, \ourmodel{} yields up to 8$\times$ speedup. In low-resource settings, with any extra costs, the ``Reuse'' module alone delivers 3–5$\times$ acceleration. 

\textbf{Internal Validity} relates to the hyperparameter-related threats and hardware environment. We conducted a focused analysis of the most influential base threshold $k$ in our discussion section. This examination specifically elucidates base threshold $k$'s impact on \ourmodel{}'s performance. Precise speedup numbers can be influenced by the specific hardware (GPUs, CPU) and software libraries used. While we report our setup, results might vary on different configurations.

\textbf{Construct Validity} relates to the metrics used in our evaluations. We use Pass@K for functional correctness and Tokens/s / Speedup for efficiency, which are standard in the field. However, Pass@K relies on test case execution and may not capture all aspects of code quality (e.g., readability, maintainability). Efficiency metrics, while objective, might not perfectly correlate with perceived developer productivity.

\section{Conclusion}

In this work, we introduce \ourmodel{}, an efficient acceleration method for LLM-based code editing. It employs a novel reuse-generate iterative paradigm via edit-oriented speculative decoding. \ourmodel{} reuses code segments from code to be edited and locates potential editing positions to efficiently generate editing content through the edit-optimized draft model enhanced and the entropy-aware dynamic verification mechanism. Experimental results on a wide range of editing benchmarks show that \ourmodel{} outperforms existing baselines in acceleration for code editing tasks, while maintaining\textemdash and in some cases improving\textemdash the quality of edited code.

\section*{Acknowledgement}
This research is supported by the National Natural Science Foundation of China Grants Nos. 62302021, 62272445, 62332001, CCF-Huawei Populus Grove Fund CCF-HuaweiSE202402, and the Fundamental Research Funds for the Central Universities (Grant No. JK2024-28).

\balance
\bibliography{ref}

\end{document}